\documentclass[11pt,a4paper]{article}
\pdfoutput=1 % if your are submitting a pdflatex (i.e. if you have
\usepackage{jheppub} % for details on the use of the package, please
\usepackage[T1]{fontenc} % if needed
\usepackage{slashed}
\usepackage{color}
\usepackage{graphicx}
\usepackage{amsmath}
\usepackage{amsfonts}
\usepackage{amssymb}
\usepackage{amsmath}
\usepackage{amssymb}
\usepackage{wasysym}
\usepackage{bm}
\usepackage{latexsym}
\usepackage{stmaryrd}
\usepackage{textcomp}
\usepackage{verbatim}
\usepackage{slashed}
\usepackage{array}

\newcommand{\of}[1]{\left(#1\right)}

\newcommand{\bigS}{\mathbb{S}}

\newcommand{\be}{\begin{equation}}
\newcommand{\ee}{\end{equation}}
\newcommand{\br}{\begin{eqnarray}}
\newcommand{\bea}{\begin{eqnarray}}
\newcommand{\eea}{\end{eqnarray}}
\newcommand{\er}{\end{eqnarray}}
\newcommand{\ba}{\begin{array}}
\newcommand{\ea}{\end{array}}
\newcommand{\bi}{\begin{itemize}}
\newcommand{\ei}{\end{itemize}}
\newcommand{\bn}{\begin{enumerate}}
\newcommand{\en}{\end{enumerate}}
\newcommand{\bc}{\begin{center}}
\newcommand{\ec}{\end{center}}

\newcommand{\Eq}[1]{Eq.~(\ref{#1})}

\newcommand{\beq}{\begin{equation}}
\newcommand{\eeq}{\end{equation}}

\newcommand{\gsim}{\lower.7ex\hbox{$\;\stackrel{\textstyle>}{\sim}\;$}}
\newcommand{\lsim}{\lower.7ex\hbox{$\;\stackrel{\textstyle<}{\sim}\;$}}

\title{Two component dark matter with multi-Higgs portals}

\author[a,1]{Ligong Bian,\note{Corresponding author.}}
\author[a,b]{Tianjun Li,}
\author[a]{Jing Shu,}
\author[a]{and Xiao-Chuan Wang,}

\affiliation[a]{State Key Laboratory of Theoretical Physics and Kavli Institute for Theoretical Physics China, Institute of Theoretical Physics, Chinese Academy of Sciences, Beijing 100190, P. R. China}
\affiliation[b]{School of Physical Electronics, University of Electronic Science and Technology of China, 
Chengdu 610054, P. R. China}

\emailAdd{lgb@itp.ac.cn}
\emailAdd{tli@itp.ac.cn}
\emailAdd{jshu@itp.ac.cn}
\emailAdd{xcwang@itp.ac.cn}
\date{\today}

\abstract{

With the assistance of 
two extra groups, i.e., an extra hidden gauge group
$SU(2)_D$ and a global $U(1)$ group,
we propose a two component dark matter (DM) model.
After the symmetry $SU(2)_D\times U(1)$ being broken, 
we obtain both the vector and scalar DM candidates. 
The two DM candidates communicate with the standard model (SM) via three Higgs as
multi-Higgs portals. The three Higgs are mixing states of the SM Higgs, the Higgs of the hidden sector 
and real part of a supplement complex
scalar singlet. We study relic density
and direct detection of DM in three scenarios. The resonance behaviors and interplay between the two 
component DM candidates are represented 
through investigating of the relic density in the parameter spaces of the two DMs masses. 
The electroweak precision parameters 
constrains the two Higgs portals couplings ($\lambda_m$ and $\delta_2$). 
The relevant vacuum stability and naturalness problem in the 
parameter space of $\lambda_m$ and $\delta_2$ are studied as well.
The model could alleviate these two problems in some parameter spaces under the constraints of
electroweak precision observables and Higgs indirect search. 
}

\begin{document}
\maketitle
\flushbottom

\section{ Introduction  }

The discovery of Higgs particle with mass around 126 GeV at Large hadron collider (LHC)~{\cite{Aad:2012tfa,Chatrchyan:2012ufa}, 
roughly consistent with the SM predictions, seems complete the SM. However, the naturalness problem is still unsolved.
To alleviate the problem, as well known and extensively studied, 
more bosonic fields are needed~\cite{Bian:2013wna,Chakraborty:2012rb,Grzadkowski:2009mj,Karahan:2014ola,Antipin:2013exa}. 
Recently, it has been noticed that the Higgs field-strength renormalization could be enhanced by 
these new boson fields~\cite{Henning:2014gca,Craig:2013xia,Englert:2013tya}.
And that after the Higgs field normalization, the Higgs coupling's modification could 
be detected indirectly~\cite{Farina:2013ssa,Craig:2013xia}.
Thus, it shed light on testing the mechanisms which alleviates the naturalness problem.
Secondly, the SM has to be extended in order to accommodate the cold dark matter (CDM) 
and baryon number 
asymmetry of the universe (BAU).
In order that the BAU not being washed out after its generation, the strong first order electroweak phase transition (SFOEWPT) is necessary, which is out  
of the capacity of the SM and new scalar fields are needed~\cite{Noble:2007kk,Damgaard:2013kva,Profumo:2014opa}\footnote{
In fact, additional heavy fermions could also make the SFOEWPT feasible, see~\cite{Davoudiasl:2012tu,Carena:2004ha}.}}.
At the same time, the way to test the mechanism alleviating the naturalness problem could also been 
used to test SFOEWPT, which is just one of the preliminary aims of the International Linear Collider (ILC)~\cite{Henning:2014gca}.
Thirdly, vacuum instability could be rescued by introducing
 vector and scalar 
fields~\cite{Gonderinger:2009jp,Drozd:2011aa,Baek:2012se,Baek:2013qwa,Gabrielli:2013hma,Bian:2013wna,Hambye:2013dgv}, and thus
making the inflation with not very small top quark mass possible to coincide with the BICEP and Plank data~\cite{Ko:2014eia,Haba:2014zda}.

All above arguments require us to extend the SM with new bosonic freedoms, 
which we choose vector and scalar fields. 
There are many studies on the multi-component DM scenarios, see~\cite{Duda:2001ae,Duda:2002hf,Profumo:2009tb,Gao:2010pg,Feldman:2010wy,Baer:2011hx,Aoki:2012ub,Chialva:2012rq,Bhattacharya:2013hva,Bian:2013wna,Kajiyama:2013rla,Esch:2014jpa}, as well as
the dynamical DM scenario~\cite{Dienes:2011ja} whose phenomenological consequences are 
often quite distinct and can be applied to a much broader variety of multi-component 
DM scenarios~\cite{Dienes:2012yz}.
The primary purpose of this work is to explain the CDM relic density, 
where the two different DM components interact with each other besides
with the three Higgses, thus affects the evolution of the DM components number densities through 
the coupled Boltzman equations~\cite{Bian:2013wna,Belanger:2011ww}.  

For the method to introduce vector dark matter through the effective 
Lagarangian method yielding strongly constraints on the parameters space from the unitarity constraints~\cite{Lebedev:2011iq,Griest:1989wd,Bian:2013wna}, we consider the scenario
in which the vector dark matter fields ($V^\mu$) respects an extra non-Abelian gauge
symmetry $SU(2)_D$. After the $SU(2)_D$ being broken via the 
the complex doublet($\phi$), one $SO(3)$ symmetry is induced
which $V^\mu$ respects to, thus making $V^\mu$ stable.\footnote{The hidden gauge theory has also been used to
study self-interacting dark matter~\cite{Hochberg:2014kqa,Hochberg:2014dra},  inflation~\cite{Yamanaka:2014pva}, and the model with a kinetic mixing portal between the
gauge boson DM and SM~\cite{Davoudiasl:2013jma}.} 
One more complex scalar singlet, $\bigS$, is supplemented to the model.
After the global $U(1)$ symmetry, i.e.,  $\bigS\rightarrow e^{i\alpha}\bigS$ respected by $V\of{H,\phi,\bigS}$, 
being broken spontaneously and softly, the real part of $\bigS$ ($S$) gets a vacuum expectation value(VEV), and the imaginary part of $\bigS$ ($A$) respects the reduced $Z_2$ symmetry, which makes
the other DM candidate.
After the breaking of the $SU(2)_D \times U(1)$, 
the Higgs field of the SM mixed with two scalar fields, the real parts of $\phi$ ($\eta'$) and $S$.
Since the two component DMs
interact with the SM particles and each other through the three Higgs fields, 
one could expect three resonance enhancement effects.
And three 
significant enlargement of the magnitude of the annihilation cross sections, of the DM to DM and the DM to SM particles,
appears.
The magnitudes of the relic density around the three resonances decrease, since which 
is robustly inversely proportional to the annihilation cross sections of the DM to SM particles.
The magnitudes of annihilation cross sections of scalar DM fields (vector DM fields) to the SM particles, 
around the three Higgs field masses, are enlarged due to
the $t$,$u$ and seagull channels of $AA(VV)\rightarrow h_ih_i$\footnote{Hereafter, the Lorenz index $\mu$ of the vector field $V^\mu$ will be
 hided for simplicity in some cases.}.  
The spin-independent (SI) DM-nucleon scattering cross sections are determined by $t$-channel interactions between 
the nucleon and two component DMs through the exchangeing of three Higgs fields. The model could be distinguished from the model with no interaction between the two DMs, since
 the opening of the channel $AA\rightarrow VV$($VV\rightarrow AA$) does affect 
 the evolution of DM number density, and thus the magnitude of relic density. This kind of effects 
 are explored and illustrated in DM relic density analysis section of this work.  
The Higgs indirect search at the LHC requires small mixing between $h$ and $\eta'$ ($S$). 
The electroweak precision observable experiment imposes stringent constraints on our parameter spaces. 
The vacuum stability of the model could be improved under the above two considerations. 
We explored the way to alleviate the naturalness problem and its' indirect
search as well.

The paper is organized as follows. We construct the model and explore relic density and direct detection of the DMs. 
After which, we investigate the Higgs indirect search and electroweak precision constraints on the model.
And then the stability of the model are studied. We also consider the possibility to alleviate naturalness problem 
and the way to trace the footprint of which.  At last, we conclude this work with discussions and conclusion.
   
\section{The Model}

To construct the model which includes the stable vector and scalar fields, we introduce these
two fields as follows.  
The vector field $V'^\mu$ is introduced as the gauged field of the $SU(2)_D$ symmetry, which couples to the SM through 
a doublet, $\phi$, which is a singlet of the SM but charged under the $SU(2)_D$. 
Here, one should note that the 
mixing between $V'^\mu$ and the SM gauge bosons through kinetic mixing is absent for the non-abelian character of the $SU(2)_D$.
We supplement one more complex singlet $\bigS$. And one more global $U(1)$ symmetry is required 
besides the $SU(2)_D$ and the SM group.
The complex singlet, which interacts with
the SM and the $SU(2)_D$ through the Higgs portal and $\phi$ portal terms in the scalar potential, transforms trivially under the SM and $SU(2)_D$ gauge group.

The Lagrangian of our model is 
 \begin{eqnarray}
{\cal L}&=& {\cal L}^{SM} -\frac{1}{4} F'^{\mu\nu} \cdot F'_{\mu \nu}+(D_\mu \phi)^\dagger (D^\mu \phi) -\frac{\mu^2_\phi}{2} \phi^\dagger \phi -\frac{\lambda_\phi}{4} (\phi^\dagger \phi)^2 \, \nonumber\\
&&+\frac{b_{2}}{2}|\bigS|^{2}
+\frac{d_{2}}{4}|\bigS|^{4}+\left(\frac{1}{4} b_{1} e^{i\phi_{b1}}\bigS^{2}+ a_{1} e^{i\phi_{a1}}\bigS+\mbox{c.c.}\right)\, \nonumber\\
&&+V_{H,\phi,\bigS}\; ,
\label{inputlagr}
\end{eqnarray}
with
\begin{eqnarray}
V_{H,\phi,\bigS}=V_{H,\phi}+V_{H,S}+V_{\phi,S}\, ,
\end{eqnarray}
and
\begin{eqnarray}
V_{H,\phi}&=&\lambda_{m}\phi^{\dagger}\phi H^{\dagger}H\, ,\\
V_{H,S}&=&\frac{\delta_{2}}{2}H^{\dagger}H|\bigS|^{2}\, ,\\
V_{\phi,S}&=&\frac{\delta_{1}}{2}\phi^{\dagger}\phi|\bigS|^{2}\; ,
\end{eqnarray}
where $D^\mu \phi=\partial^\mu \phi - i\frac{g_\phi}{2} \tau \cdot V'^\mu$.  The hidden gauge coupling $g_\phi < 4\pi$, 
required by the unitarity bound, need to be hold for any thermal particle whose relic density arises 
from the freeze-out of its annihilation~\cite{Frampton:1989fu,Liu:2008bh,Lebedev:2011iq}. 
In particular, $b_1$ and $a_1$ terms break the global $U(1)$ symmetry explicitly.
In the SM Lagrangian, the Higgs potential notations 
are defined as: ${\cal L}^{SM} \owns -(m^2/2) H^\dagger H -\lambda (H^\dagger H)^2$ with $H=(0,v+h)/\sqrt{2}$, where 
$v$ is the VEV of the Higgs field.

After the $SU(2)_D$ being spontaneously broken, both $\phi$ and the singlet $\bigS$ get the VEVs:
 \begin{eqnarray}
&&\phi=(0,~ \frac{v_{\phi}+\eta^{\prime}}{\sqrt{2}})\; ,\\
&&\bigS=\frac{1}{\sqrt{2}}(v_s+S+iA)\; .
\end{eqnarray}
  So the Eq.~(\ref{inputlagr}) in the unitary gauge recast as,
\begin{eqnarray}
{\cal L}&=&{\cal L}_{SM}-\frac{1}{4} F_{\mu \nu} \cdot F^{\mu \nu} 
+\frac{1}{8} (g_\phi v_\phi)^2 V_\mu \cdot V^\mu\nonumber\\
&& +\frac{1}{8} g_\phi^2 V_\mu \cdot V^\mu \eta'^2
+\frac{1}{4} g_\phi^2 v_\phi V_\mu \cdot V^\mu \eta' +\frac{1}{2}(\partial_\mu \eta')^2
\nonumber\\
&&\,\nonumber\\
&&+V_{h,\eta',S,A}~,~
\label{lagrzerov}
\end{eqnarray}
here $V^{\mu}=U V'^{\mu} U^{-1}-\frac{i}{g} [\partial_\mu U] U^{-1}$ with $U={\rm exp}(-i\tau \cdot \xi/v_\phi)$.
The vector DM mass is given by $m_{V}=g_\phi v_\phi/2$,
and the tree-level potential $V_0\of{h,\eta^\prime,S,A}$ recast as,
\begin{eqnarray}
\label{eq:potential_tree2}
V_0\of{h,\eta^\prime,S,A} &=& \frac{m^2}{4}(h+v)^2+\frac{\mu_\phi^2}{4} (\eta'+v_\phi)^2 - \frac{\lambda_\phi}{16} (\eta'+v_\phi)^4 + \frac{1}{4}\lambda (h+v)^4\nonumber\\
&&
+\frac{1}{4}\lambda_m (h+v)^2 (\eta^\prime+v_\phi)^2+ \frac{1}{8}(\delta_2 (h+v)^2+\delta_1 (\eta^\prime+v_\phi)^2)\of{(S+v_s)^2+A^2} 
\nonumber\\
&&
+\frac{1}{4}\of{b_2-b_1}(S+v_s)^2+ \frac{1}{4}\of{b_2+b_1}A^2 - \sqrt{2}a_1 (S+v_s)
+ \frac{d_2}{8}(S+v_s)^2A^2  \nonumber\\
&&+ \frac{d_2}{16}\of{(S+v_s)^4 + A^4}\, .
\end{eqnarray}
Here, we would like to mention that,  $\eta'$ lives in the fundamental representation of $SU(2)_D$, and
 displays a custodial symmetry $SO(3)$ in the $V^\mu_{1,2,3}$ component space, which makes 
 three $V^{\mu}_i$ components degenerate in mass and thus stable~\cite{Hambye:2008bq}. 
The explicit $Z_2$-breaking term is proportional to $a_1$ and being introduced here to avoid the
 cosmological domain wall problem~\cite{Zeldovich:1974uw,Kibble:1976sj,Kibble:1980mv}.
After choosing $\phi_{a_{1}}=\phi_{b_{1}}=\pi$, the potential retains a $Z_2$ symmetry for $\mathrm{Im} (\bigS)$, 
thereby ensuring the stability of the particle $A$~\cite{Barger:2008jx,Gonderinger:2012rd}.  

Requiring that the potential in Eq.~(\ref{eq:potential_tree2}) has a minimum at $\langle H\rangle = h/\sqrt{2} = 0$ 
and $\langle \bigS\rangle = S+iA = 0+i\cdot 0$, the following minimization conditions 
are obtained:
\begin{equation}\label{eq:tree_min_cond}
\frac{\partial V_0}{\partial h} = 0,\ \ \frac{\partial V_0}{\partial \eta} = 0,\ \ \frac{\partial V_0}{\partial S} = 0,\ \ \frac{\partial V_0}{\partial A} = 0~,~
\end{equation}
where all derivatives are evaluated at $(h,\eta,S,A)=(0,0,0,0)$.  These minimization conditions allow the Higgs VEV $v$ and the singlet VEV $v_s$ to replace $m^2$ and $b_2$  according to
\begin{equation}\label{eq:msq_b2_replace}
\begin{aligned}
&m^2 \equiv -\frac{1}{2}(4v^2 \lambda+v_s^2 \delta_2-2v_\phi^2\lambda_m) \; ,\\
&\mu_\phi^2\equiv -\frac{1}{2}(\delta_1 v_s+2\lambda_m v^2+\lambda_\phi v_\phi^2)\; ,\\
&b_2 \equiv \frac{1}{2v_s}(4\sqrt{2}a_1-v_s(-2b_1+d_2 v_s^2+\delta_1 v_\phi^2+\delta_2 v^2))\; .\\
\end{aligned}
\end{equation}
Thus, at the minima, the  mass matrix is obtained as
\begin{equation}
\label{eq:tree_mass_matrix}
M=\left(\begin{array}{cccc}
2\lambda v^{2}&\lambda_{m}vv_{\phi}&\frac{1}{2}\delta_{2} v_s v&0\\
\lambda_{m}vv_{\phi}&\frac{1}{2}\lambda_{\phi}v_{\phi}^{2}&\frac{1}{2}\delta_{1} v_s v_{\phi}&0\\ 
\frac{1}{2}\delta_{2} v_s v&\frac{1}{2}\delta_{1} v_s v_{\phi}&\frac{2\sqrt{2}a_{1}+v_s^{3}d_{2}}{v_s}&0\\
0&0&0&\frac{a_1}{2\sqrt{2}v_s}+b_1
\end{array}\right)\, .
\end{equation}
%\begin{multline}\label{eq:tree_mass_matrix}
%\lb\begin{array}{ccc} m_h^2 & m_{hS}^2 & m_{hA}^2 \\ m_{hS}^2 & m_S^2 & m_{SA}^2 \\ m_{hA}^2 & m_{SA}^2 & m_A^2\end{array}\rb =\\
%\lb\begin{array}{ccc} \frac{1}{2}\lambda v^2 & \frac{1}{2}\del xv & 0 \\ \frac{1}{2}\del xv & \frac{1}{2}d_2x^2 + \sqrt{2}a_1/x & 0 \\ 0 & 0 & b_1 + \sqrt{2}a_1/x\end{array}\rb 
%\end{multline}
And in the basis of $(h,\ \eta^{\prime},\ S)$, we have
\begin{eqnarray}
\label{eigen}
M=\left(\begin{array}{ccc}
2\lambda v^{2}&\lambda_{m}vv_{\phi}&\frac{1}{2}\delta_{2} v_s v\\
\lambda_{m}vv_{\phi}&\frac{1}{2}\lambda_{\phi}v_{\phi}^{2}&\frac{1}{2}\delta_{1} v_s v_{\phi}\\ 
\frac{1}{2}\delta_{2} v_s v&\frac{1}{2}\delta_{1} v_s v_{\phi}&\frac{2\sqrt{2}a_{1}+{v_s}^{3}d_{2}}{v_s}
\end{array}\right)\; .
\end{eqnarray}
To work in the mass eigenstates, i.e., $h_{1,2,3}$, we  diagonalise the mass matrix Eq.~(\ref{eigen})  through
\begin{equation}
RM^{2}R^{T}=M_{diag}^{2},
\end{equation}
with matrix $R$ being given by
\begin{eqnarray}
\label{Rvals}
R=\left(\begin{array}{ccc}
c_{1}c_{3} & c_{3}s_{1} & s_{3}\\
c_{2}s_{1}-c_{1}s_{2}s_{3} & c_{1}c_{2}-s_{1}s_{2}s_{3} & c_{3}s_{2}\\
s_{1}s_{2}-c_{1}c_{2}s_{3} & c_{1}s_{2}-c_{2}s_{1}s_{3} & c_{2}c_{3}
\end{array}\right),
\end{eqnarray}
$c,\,s$ and their subscripts $1,\,2,\,3$ represent 
$\cos,\,\sin$, $\theta_{12},\,\theta_{23}$, and $\theta_{13}$ individually.
\section{Dark matter analysis}
\begin{figure}[!htb]
\begin{minipage}[t]{0.4\linewidth} 
    \centering
    \includegraphics[width=0.8\linewidth]{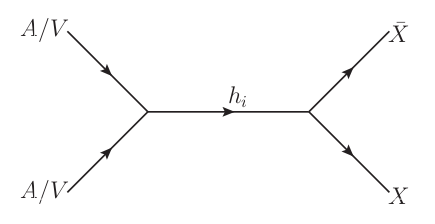} 
\end{minipage}
\hspace{0.05\textwidth}
\begin{minipage}[t]{0.5\linewidth} 
    \centering
    \includegraphics[width=0.6\linewidth]{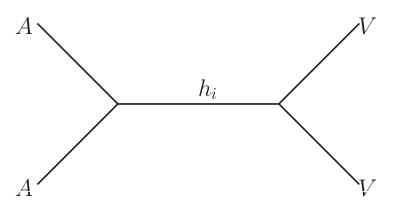} 
\end{minipage}
~\\
\begin{minipage}[t]{0.45\textwidth} 
    \centering
    \includegraphics[width=0.45\textwidth]{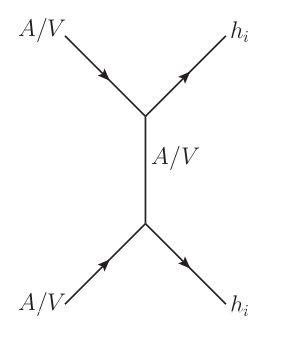} 
\end{minipage}
\hspace{0.05\textwidth}
\begin{minipage}[t]{0.45\textwidth} 
    \centering
    \includegraphics[width=0.45\textwidth]{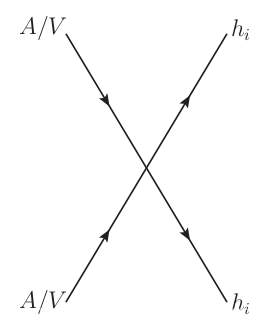} 
\end{minipage}
    \caption{Feynman diagrams of annihilation channels.}
    \label{fig:feyndia}
\end{figure}

In our model, we have two component DMs. Thus, to present the novelty of the model properly,
we use the coupled boltzman equations explored in~\cite{Bian:2013wna}, with annihilation channels being depicted in
Fig.~\ref{fig:feyndia}, and details of annihilation cross sections are listed in section~\ref{annh}.
The relic density of each component DM could be obtained through 
$\Omega_{A,V}h^{2}=2.755\times10^{8}\frac{M_{A,V}}{GeV}Y_{A,V}(T_{0})$~\cite{Edsjo:1997bg,Biswas:2013nn} after we calculated $Y_{A,V}(T_{0})$ numerically, and the total relic density is the sum of the two component DMs, $\Omega h^{2}=\Omega_{A}h^{2}+\Omega_{V}h^{2}$. As for the numerical calculations of $Y_{A,V}(T_{0})$, we refer to Eqs.~(\ref{eq:boltzAA}), (\ref{eq:boltzVA}),
(\ref{eq:boltzVV}), and (\ref{eq:boltzAV}).

In order to investigate how does the interplay between the two DM components affect the evolution
of the DM abundance and the
dependence of the DM relic density on each
parameters, we scan the parameter spaces according to the following three groups:
 \begin{enumerate}
   \item {Scan in the $M_{A} - M_{V}$ plane for the case of $M_{A}>M_{V}$, with the other parameters being
given in Table~\ref{tab:mAmV}.}
   \item {Scan in the $M_{A} - M_{V}$ plane for the case of $M_{A}<M_{V}$, with the other parameters being
given in Table~\ref{tab:mAmV}.}
   \item {Scan in the $\lambda_{m} - \delta_2$ plane, with DM masses being fixed as: $M_V$=120 GeV, $M_{A}$=150 GeV, and the other
   parameters are given in Table~\ref{tab:VEV}.}
 \end{enumerate}

\begin{table}[!htp]
\begin{tabular}{|c|c|c|c|c|c|c|c|c|c|c|}
\hline 
$v$ & $v_{\phi}$ & $v_{s}$ & $\lambda$ & $\lambda_{\phi}$ & 
$d_2$ & $\delta_{1}$ & $\delta_{2}$ & $\lambda_{m}$ & $g_{\phi}$\tabularnewline
\hline 
$246$ & $738$ & $123$ & $0.515$ & $0.2$ & 
$0.2$ & $0.02$ & $0.04$ & $0.01$ & $0.04$\tabularnewline
\hline 
\end{tabular}
\caption{The input parameters for the analyses of 
the cases of $m_A>m_V$ and $m_A<m_V$.}
\label{tab:mAmV}
\end{table}

\begin{table}[!htp]
\begin{tabular}{|c|c|c|c|c|c|c|c|}
\hline 
$v$ & $v_{\phi}$ & $v_{s}$&$\lambda$ & $\lambda_{\phi}$ & $d_{2}$ & $\delta_1$\tabularnewline
\hline 
$246$ & $738$ & $123$&$0.515$ & $0.2$ & $0.21$ & $0.02$\tabularnewline
\hline 
\end{tabular}
\caption{The VEVs of three scalar fields and other couplings.}
\label{tab:VEV}
\end{table}

\subsection{Relic density analysis}
\label{sec:rda}

Set free parameters as in Table~\ref{tab:mAmV}, the three Higgs masses are obtained as 
$m_{h_1}=124.8,~\,m_{h_2}=233.8$, and $m_{h_3}=71.1$~GeV in order.

\begin{enumerate} 
\item{{\bf The first group with $m_A>m_V$}}
~\\
\begin{figure}[!htb]
\begin{minipage}[t]{0.45\textwidth} 
    \centering
    \includegraphics[scale=0.65]{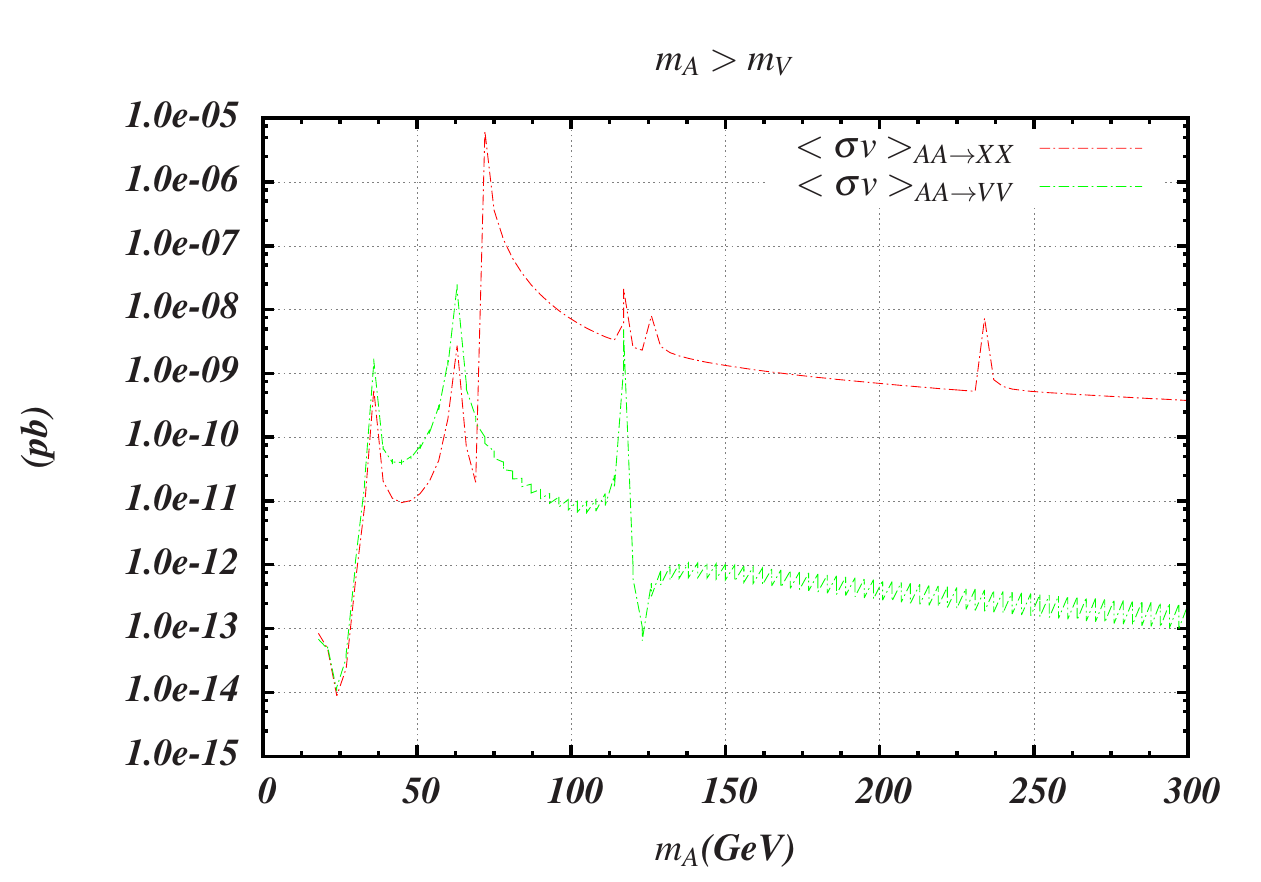} 
\end{minipage}
\hspace{0.5cm}
\begin{minipage}[t]{0.45\textwidth} 
    \centering
    \includegraphics[scale=0.65]{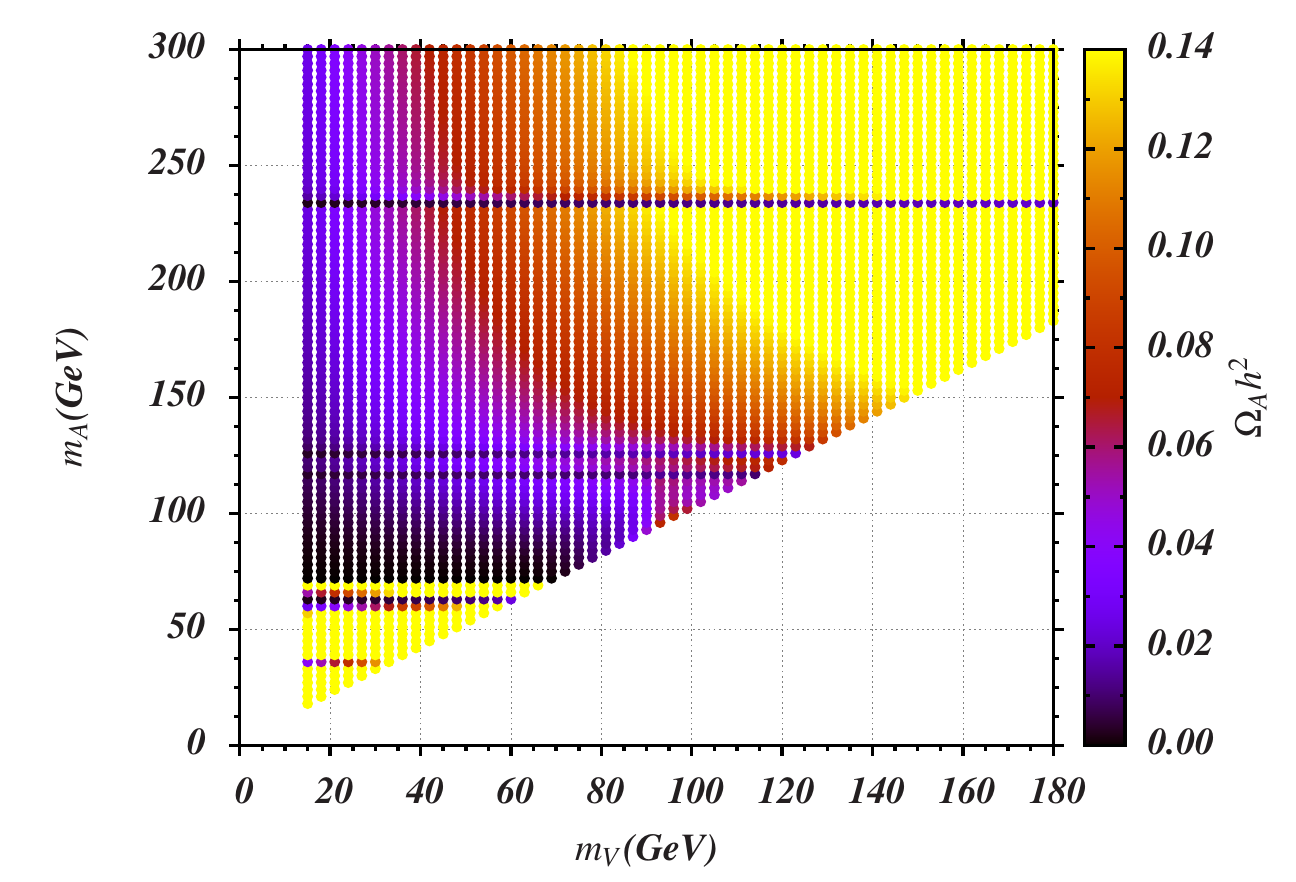} 
\end{minipage}
    \caption{Left: Plots for the cross sections of 
    channels $AA\to~XX$ and $AA\to~VV$ in units of $0.3894\times 10^9 ~pb$; Right: 
    The corresponding relic density $\Omega_{A}h^2$.}
    \label{fig:mAmVomgA}
\end{figure}
 
 \begin{figure}[!h]
\begin{minipage}[t]{0.45\textwidth} 
    \centering
    \includegraphics[scale=0.65]{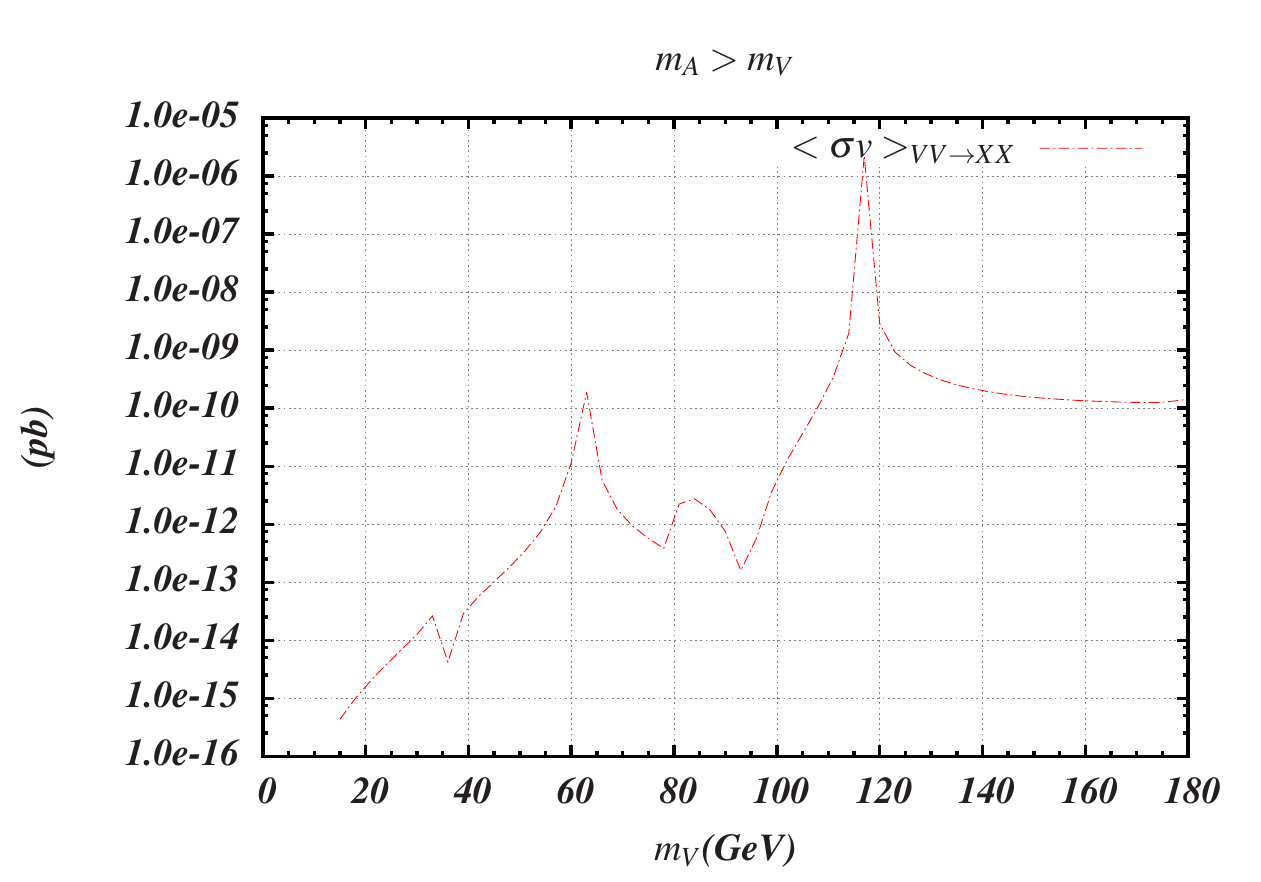} 
\end{minipage}
\hspace{0.5cm}
\begin{minipage}[t]{0.45\textwidth} 
    \centering
    \includegraphics[scale=0.65]{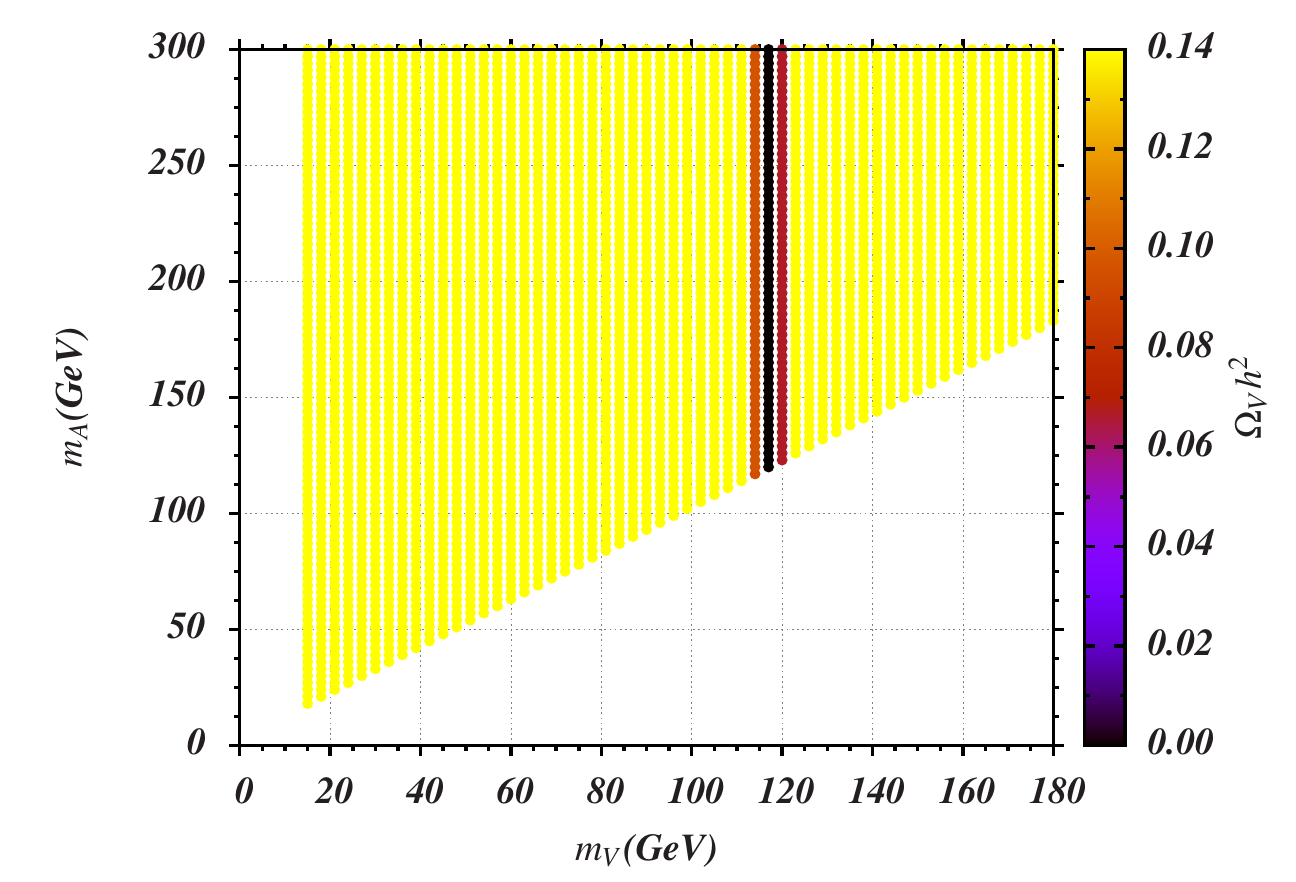} 
\end{minipage}
    \caption{Left: Plot of the cross sections of 
    channel $VV\to~XX$ in units of $0.3894\times 10^9 ~pb$; Right: Corresponding 
    relic density $\Omega_{V}h^2$ for $m_A>m_V$.}
    \label{fig:mAmVomgV2}
\end{figure}

 The annihilation channel $AA\to~VV$ 
is opened up and shown by the green line on the left panel of 
Fig.~\ref{fig:mAmVomgA}. 
The bandwidth of the green line comes from the 
change of the vector DM mass $m_V$, and the bandwidth
becomes wider and wider with the increasing of scalar 
DM mass $m_A$, since the bigger the scalar DM mass $m_A$ 
the larger viable range will be left for $m_V$.
Fig.~\ref{fig:mAmVomgA} depicts that the channel $AA\to~XX$ is the dominate one when
$m_A>$70 GeV, and the variation tendency of the relic density $\Omega_{A}h^2$ is plotted in the right panel. 
Three peaks of the annihilation cross section of 
$AA\to~VV$ are presented, at $m_A=m_{h_1}/2,\,m_{h_2}/2,\,m_{h_3}/2$, 
those are the three resonances at these mass values, 
as could be seen from Eq.~(\ref{AAVV}) as well. 
At the same time, 
we find three other peaks of the annihilation cross section of $AA\to~XX$, besides the same three peaks as that of $AA\to~VV$, 
at $m_A=m_{h_i}$
(the three masses of three Higgs),
because the new partial $t$, $u$ and seagull 
channels $AA\to~h_{i}h_{i}$ through exchanging the 
scalar DM particle themselves are opened up around these mass values. 
These peaks represent themselves on the right panel of  Fig.~\ref{fig:mAmVomgA} 
by the transition of the color along axis of $m_A$.
When there is a peak bigger than about $1\times~10^{-9}$ 
for the cross section (on the left panel), a decrease behavior
of the corresponding relic density appears on the right panel.

Last but not the least, we would like to emphase that the opening 
of the channel $AA\rightarrow VV$ does cause the decrease of the magnitude of $\Omega_A h^2$.
Because of $\Omega_A h^2\sim1/\langle\sigma v\rangle_{AA\rightarrow XX}$, when the channel 
$AA\rightarrow VV$ has been shut down, one may expect $\langle\sigma v\rangle_{AA\rightarrow XX}\sim pb$
and $\Omega_A h^2\sim 0.1$. While, the right panel of Fig.~\ref{fig:mAmVomgA}
depicts more decrease of the magnitude of $\Omega_A h^2$ around $m_A=m_{h_{1,2}}/2$, which demonstrates
the effects of the annihilation process $AA\rightarrow VV$.

For the case of $m_A>m_V$, as shown in Fig.~\ref{fig:mAmVomgV2},  
the channel of $VV\to~AA$ is closed when we analyzing $\Omega_Vh^2$ .
The cross section of the channel $AA\to~VV$ together 
with that of $VV\to~XX$ affect 
the corresponding relic density $\Omega_Vh^2$, 
and the second one dominates the value of $\Omega_Vh^2$. Only the slim peak 
caused by the resonance at about $m_V=m_{h_2}/2$ brings a sizable decreasing of $\Omega_Vh^2$. 
The viable region of $m_V$ matching the experimental value 
$\Omega h^2=0.1189$~\cite{PDG:relicdensity}($\Omega_Vh^2$) is very small as shown in the right panel. 

\item{{\bf The second group with $m_V>m_A$}}
\begin{figure}[!htb]
\begin{minipage}[t]{0.45\textwidth} 
    \centering
    \includegraphics[scale=0.6]{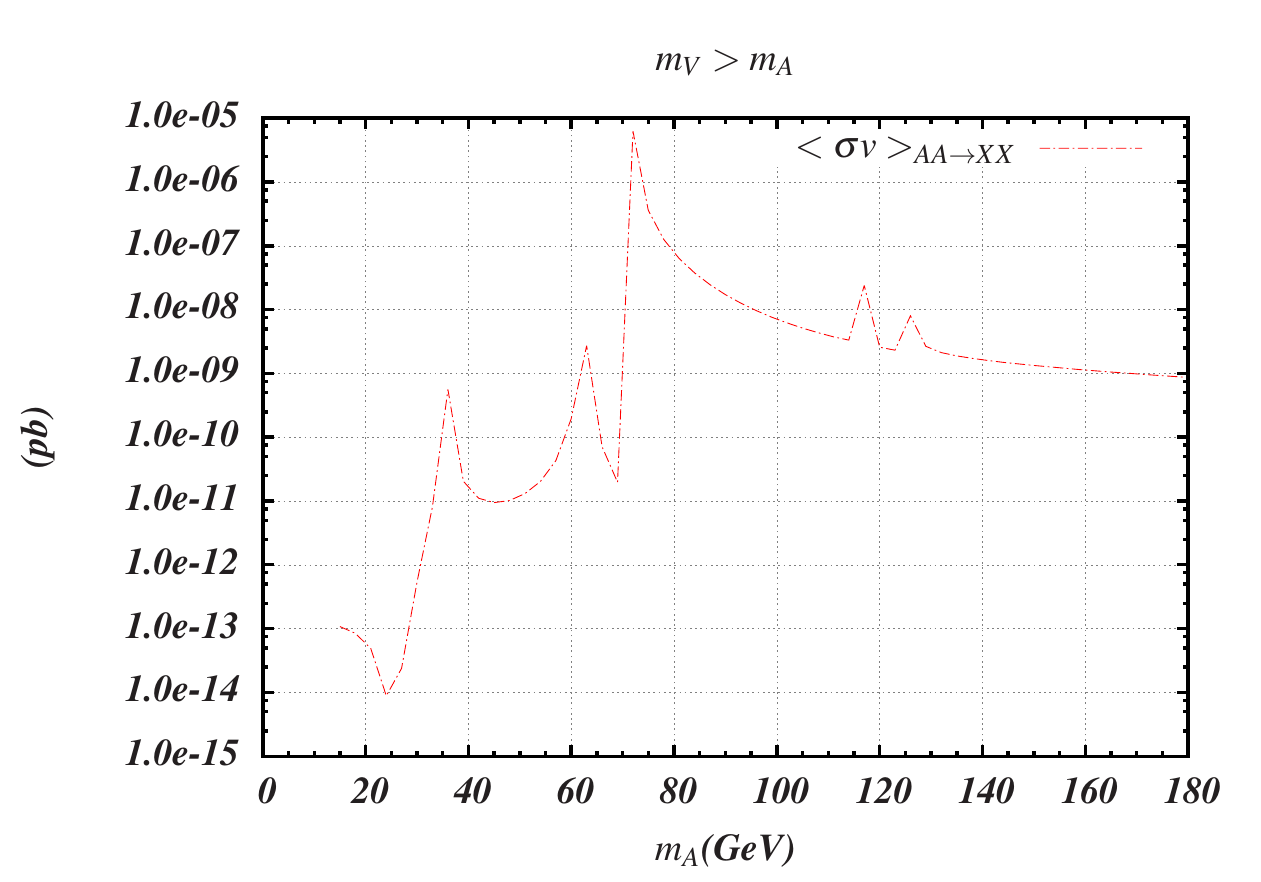} 
\end{minipage}
\hspace{0.5cm}
\begin{minipage}[t]{0.45\textwidth} 
    \centering
    \includegraphics[scale=0.6]{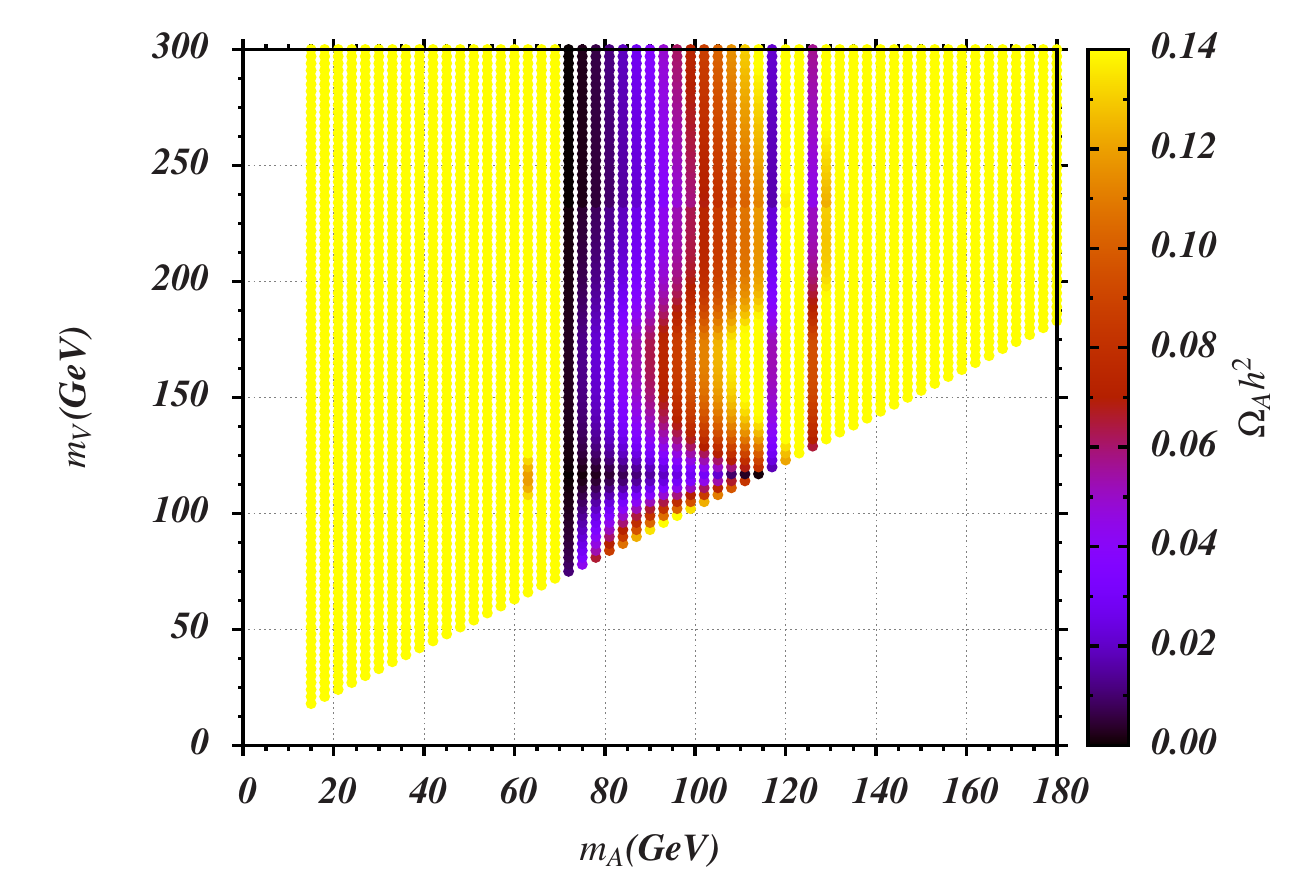} 
\end{minipage}
    \caption{ $m_V>m_A$. 
    Left: Plots for annihilation cross section of 
    channel $AA\to~XX$ in units of $0.3894\times 10^9 ~pb$; Right:  
    Relic density $\Omega_{A}h^2$.}
    \label{fig:mVmAomgA}
\end{figure}
\begin{figure}[!htb]
\begin{minipage}[t]{0.45\textwidth} 
    \centering
    \includegraphics[scale=0.6]{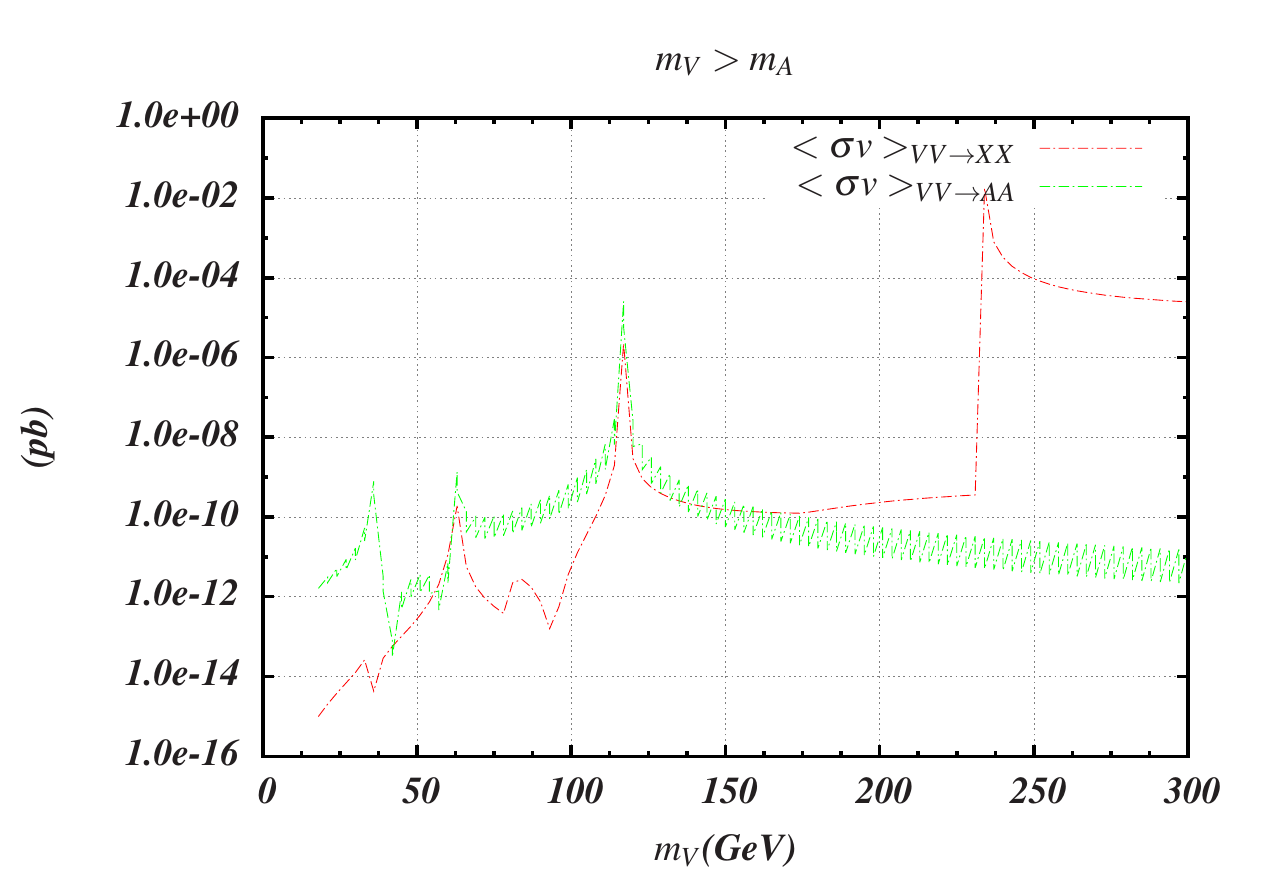} 
\end{minipage}
\hspace{0.5cm}
\begin{minipage}[t]{0.45\textwidth} 
    \centering
    \includegraphics[scale=0.6]{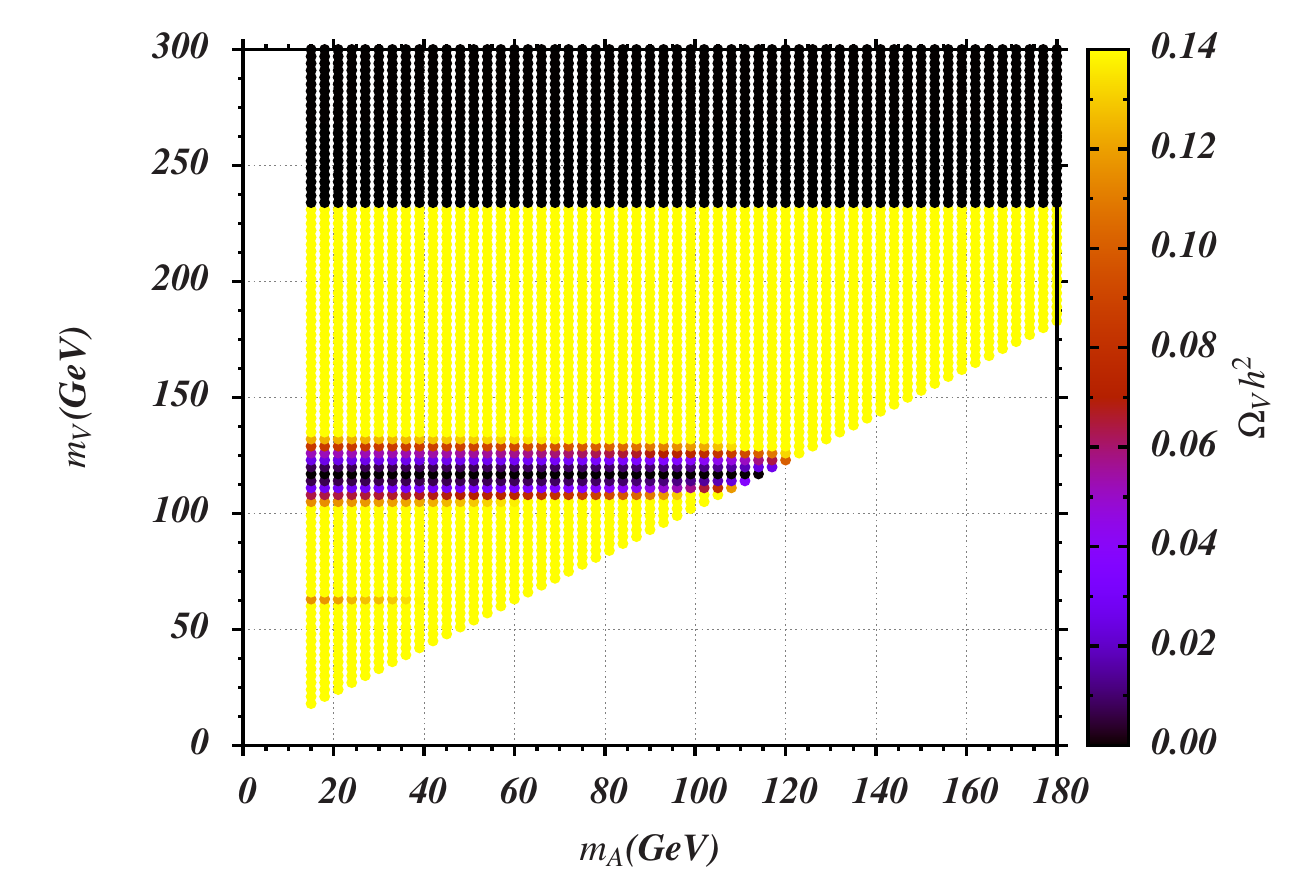} 
\end{minipage}
    \caption{$m_V>m_A$. Left: Plots for the cross sections of annihilation
    channels $VV\to~XX$ and $VV\to~AA$ in units of $0.3894\times 10^9 ~pb$; Right: The dominating 
    relic density $\Omega_{V}h^2$.}
    \label{fig:mVmAomgV}
\end{figure}

In this case, the channel $AA\to~VV$ 
is forbidden due to $m_V>m_A$, and the behavior of the cross section of
the channel $AA\to~XX$ is given
in the left panel of Fig.~\ref{fig:mVmAomgA}. With the increasing of $m_A$, 
the first peak caused by the resonance effect does not leave any trails, 
as could be seen in the magnitude of the corresponding relic density (see the right panel of Fig.~\ref{fig:mVmAomgA}), 
because the magnitude of the cross section is too small. 
The second peak is caused by the resonance effect at $m_A=m_{h_1}/2$,  
and the tinny increasing of cross section induces a small decreasing of the corresponding relic density. 
The third peak demonstrates the opening up of the partial $t$ and $u$ channels for $AA\to~h_3h_3$ and also 
brings a very big decrease of the relic density at 
about $m_A\ge~m_{h_3}$ as expected. And the fourth peak, which is caused by the resonance at $m_A=m_{h_2}/2$, brings 
a sizable decrease of the magnitude of the relic density also.
We also notice the small waves at about $m_V=160$ GeV in the right panel of Fig.~\ref{fig:mVmAomgA}.
The phenomenon could be explained by Eq.~(\ref{eq:boltzAV}), which illustrates that the behavior of the cross section $\langle\sigma v\rangle_{VV\to~AA}$ affects the calculation of $\Omega_Ah^2$ 
and the variation of the magnitude of the cross section $\langle\sigma v\rangle_{VV\to~AA}$ 
generates these small waves. The channel $VV\to~AA$ also causes a bigger decrease of the magnitude of $\Omega_Vh^2$ in comparison with the scenario in which the channel $VV\to~AA$ has been 
shut down, as shown in the right panel of Fig.~\ref{fig:mVmAomgV}, especially around $m_V\sim m_{h_1}/2(m_{h_1})$. 
Fig.~\ref{fig:mVmAomgV} is plotted to show the magnitude of $\Omega_Vh^2$
contributing from both channels $VV\to~XX$ and $VV\to~AA$.
The cross sections are small for most regions, and a big decrease of $\Omega_Vh^2$
exists in the two regions 
$m_V\sim~m_{h_2}/2$ and $m_V>m_{h_2}$. For the first decrease, both channels 
induce comparable effects. And for the second decrease, it is the channel $VV\to~XX$
that dominates the magnitude of $\Omega_Vh^2$.
Especially when $m_V>m_{h_2}$, the partial $t$, $u$ and seagull 
channels $VV\to~h_2h_2$ through exchanging 
the vector DM particle are opened up. 
These effects give rise to too big $\langle\sigma v\rangle_{VV\to~XX}$, and then the door to generate relic density 
$\Omega_Vh^2$ is almost closed. 

\begin{figure}[!htb]
\begin{minipage}[t]{0.45\textwidth} 
    \centering
    \includegraphics[scale=0.5]{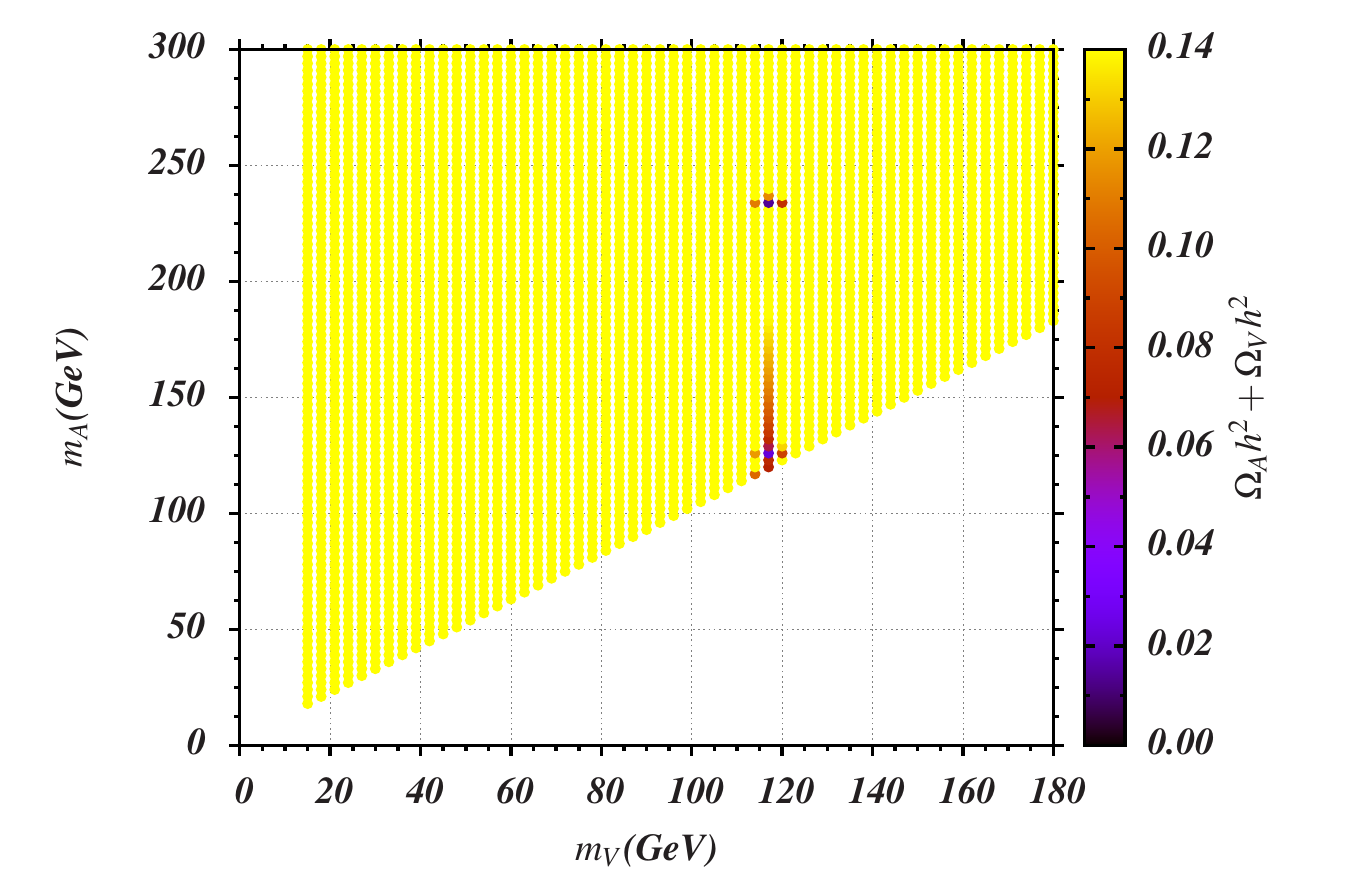} 
\end{minipage}
\hspace{0.5cm}
\begin{minipage}[t]{0.45\textwidth} 
    \centering
    \includegraphics[scale=0.5]{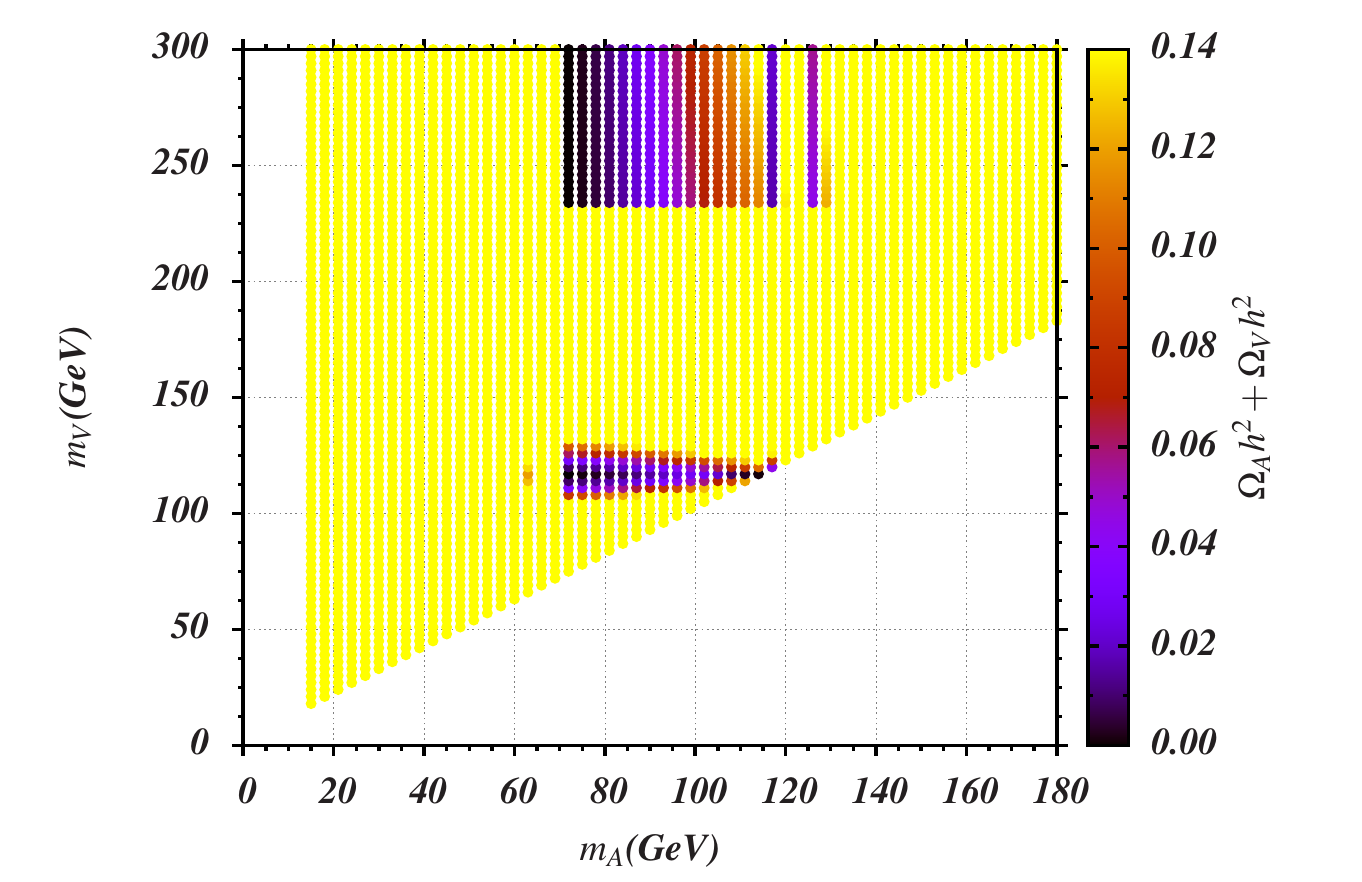} 
\end{minipage}
    \caption{Plots of the total relic density 
    $\Omega_{A}h^2+\Omega_{V}h^2$ for 
    the case of $m_A>m_V$ and $m_V>m_A$.}
    \label{fig:mAmVomgall}
\end{figure}

At last, the two panels in the Fig.~\ref{fig:mAmVomgall} presents the total relic density 
 for $m_A>m_V$ 
and $m_V>m_A$, respectively.
Under the set of parameters as in 
the Table~\ref{tab:mAmV}, the large enough magnitude of the relic density could be obtained easily.

\begin{figure}[!htb]
\begin{minipage}[t]{0.45\textwidth} 
    \centering
    \includegraphics[scale=0.5]{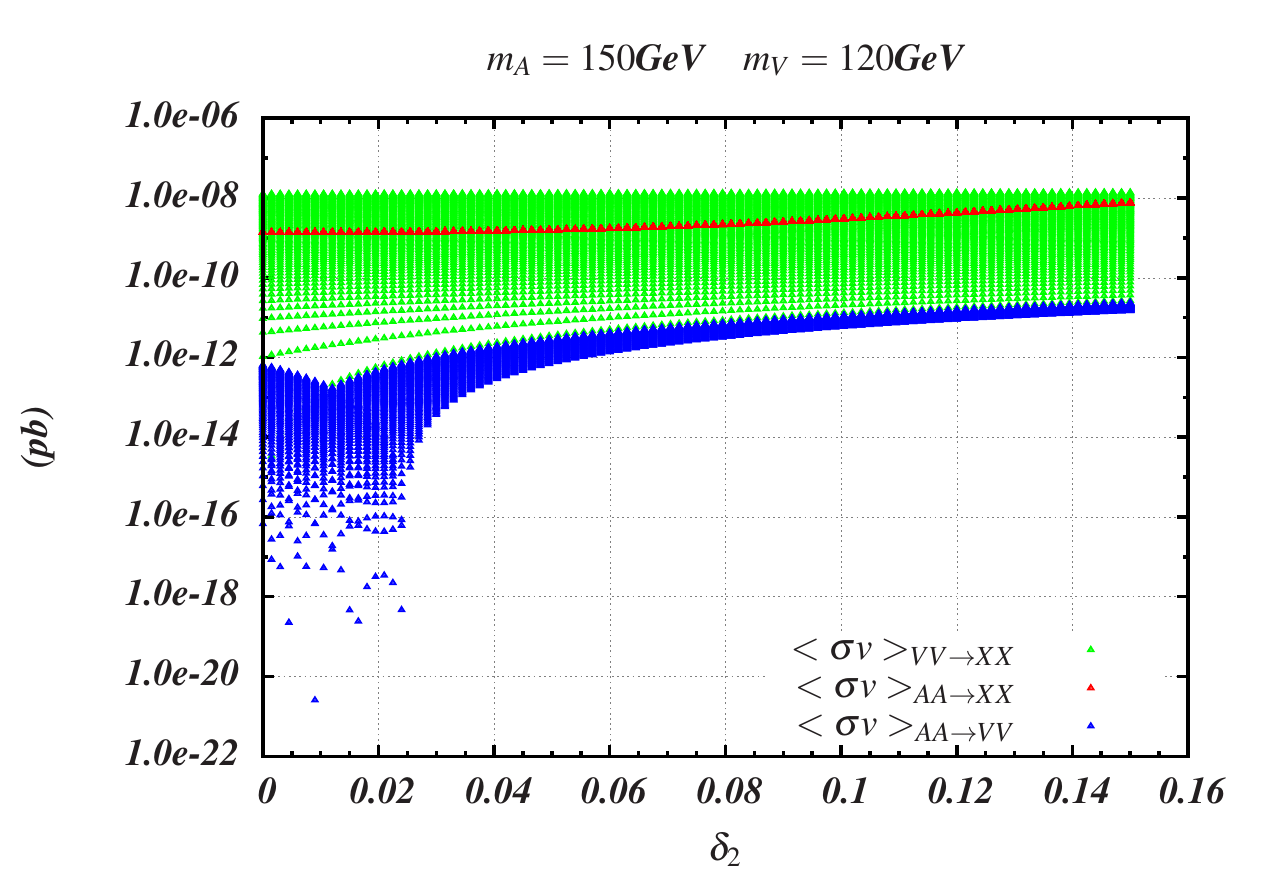} 
\end{minipage}
\hspace{0.003\textwidth}
\begin{minipage}[t]{0.45\textwidth} 
    \centering
    \includegraphics[scale=0.5]{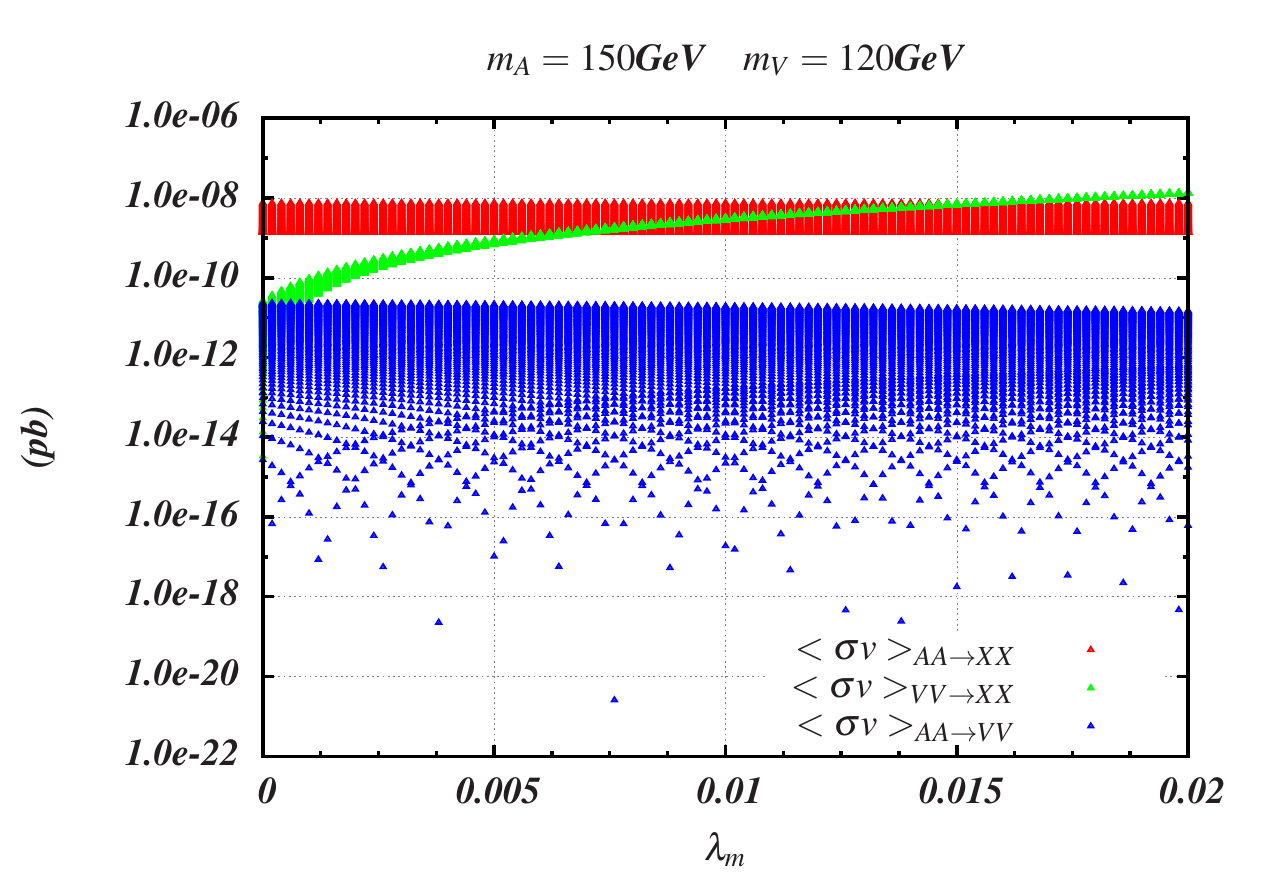} 
\end{minipage}
    \caption{Plots of $\delta_{2}$ and $\lambda_{m}$ $vs$. cross sections.}
    \label{fig:dlt2lmdmcorsec}
\end{figure}
\begin{figure}[!htb]
\begin{minipage}[t]{0.45\textwidth} 
    \centering
    \includegraphics[scale=0.6]{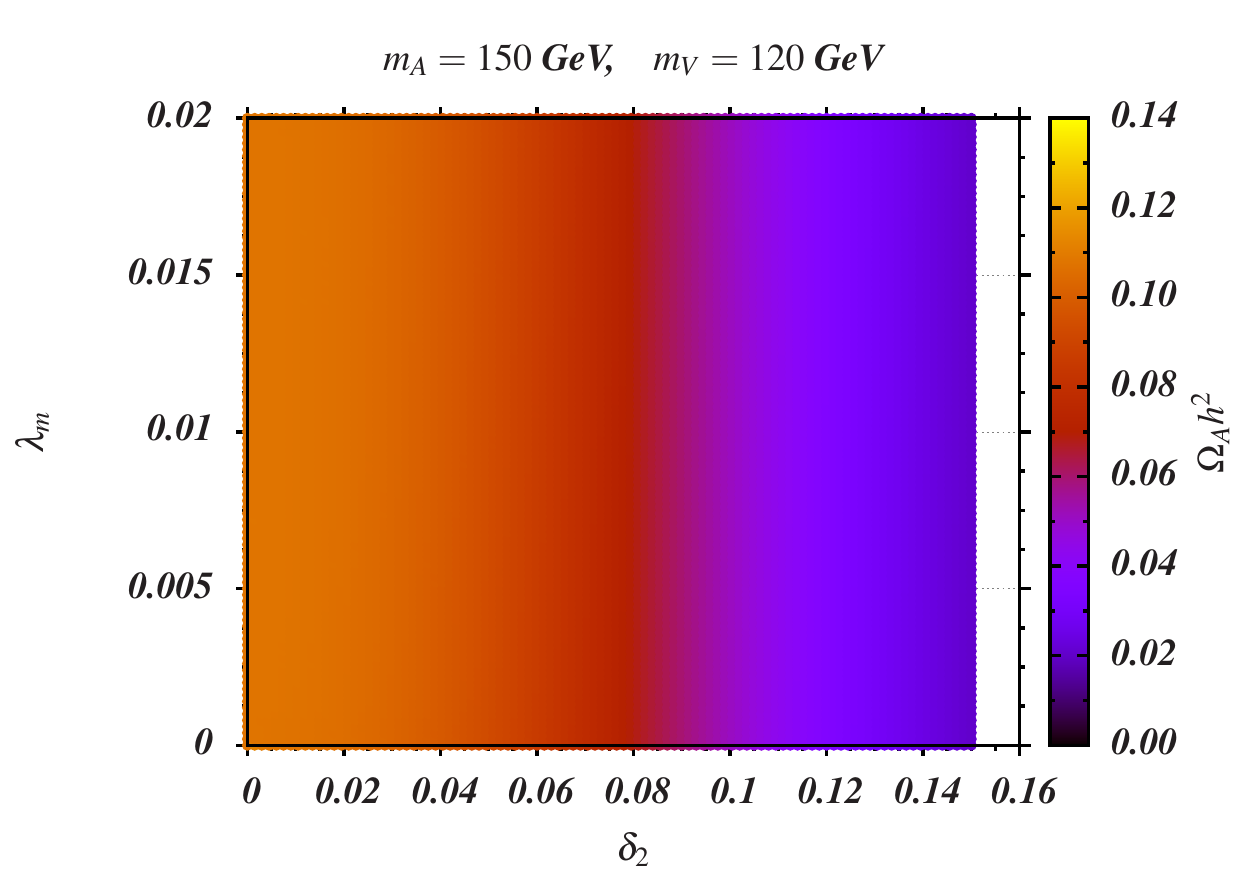} 
\end{minipage}
\hspace{0.003\textwidth}
\begin{minipage}[t]{0.45\textwidth} 
    \centering
    \includegraphics[scale=0.6]{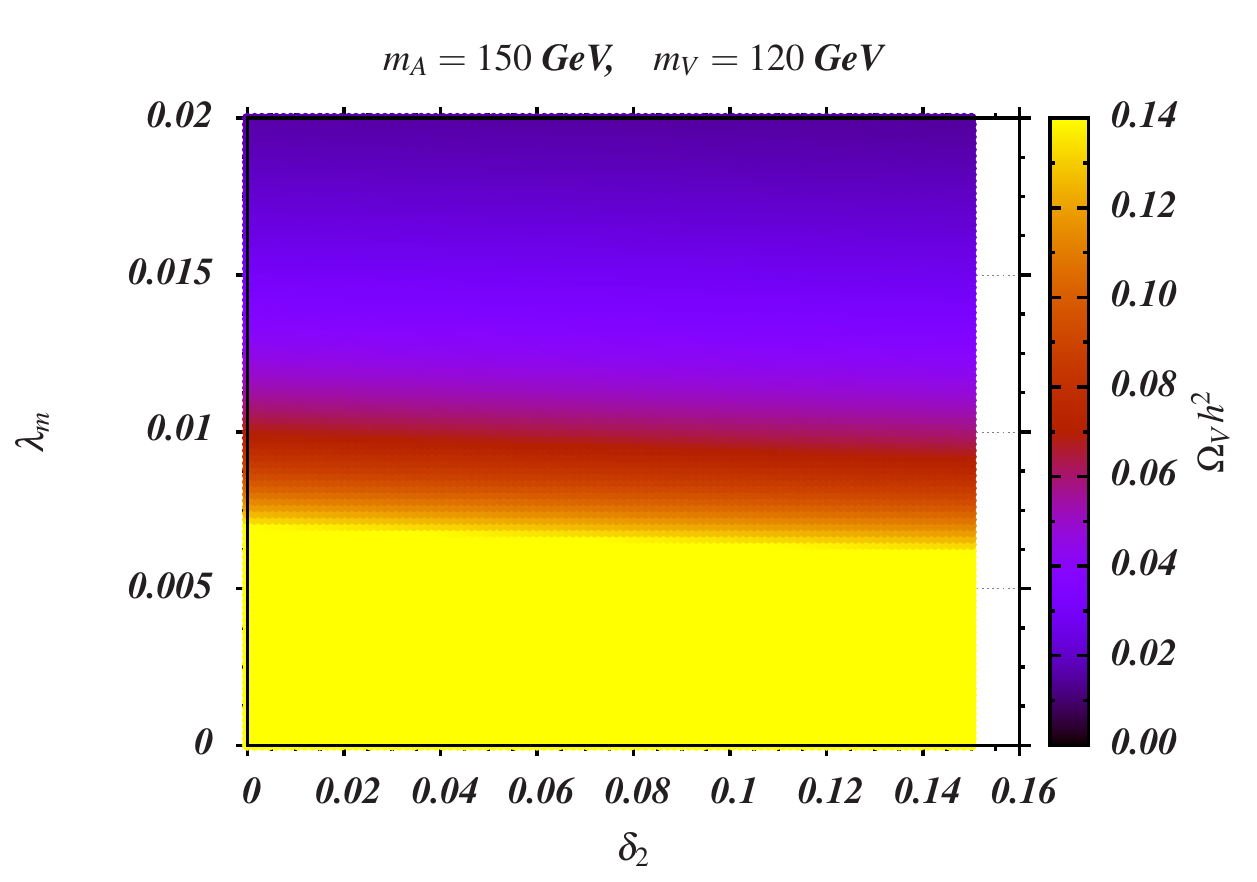} 
\end{minipage}
\hspace{0.003\textwidth}
\begin{minipage}[t]{0.9\textwidth} 
    \centering
    \includegraphics[scale=0.6]{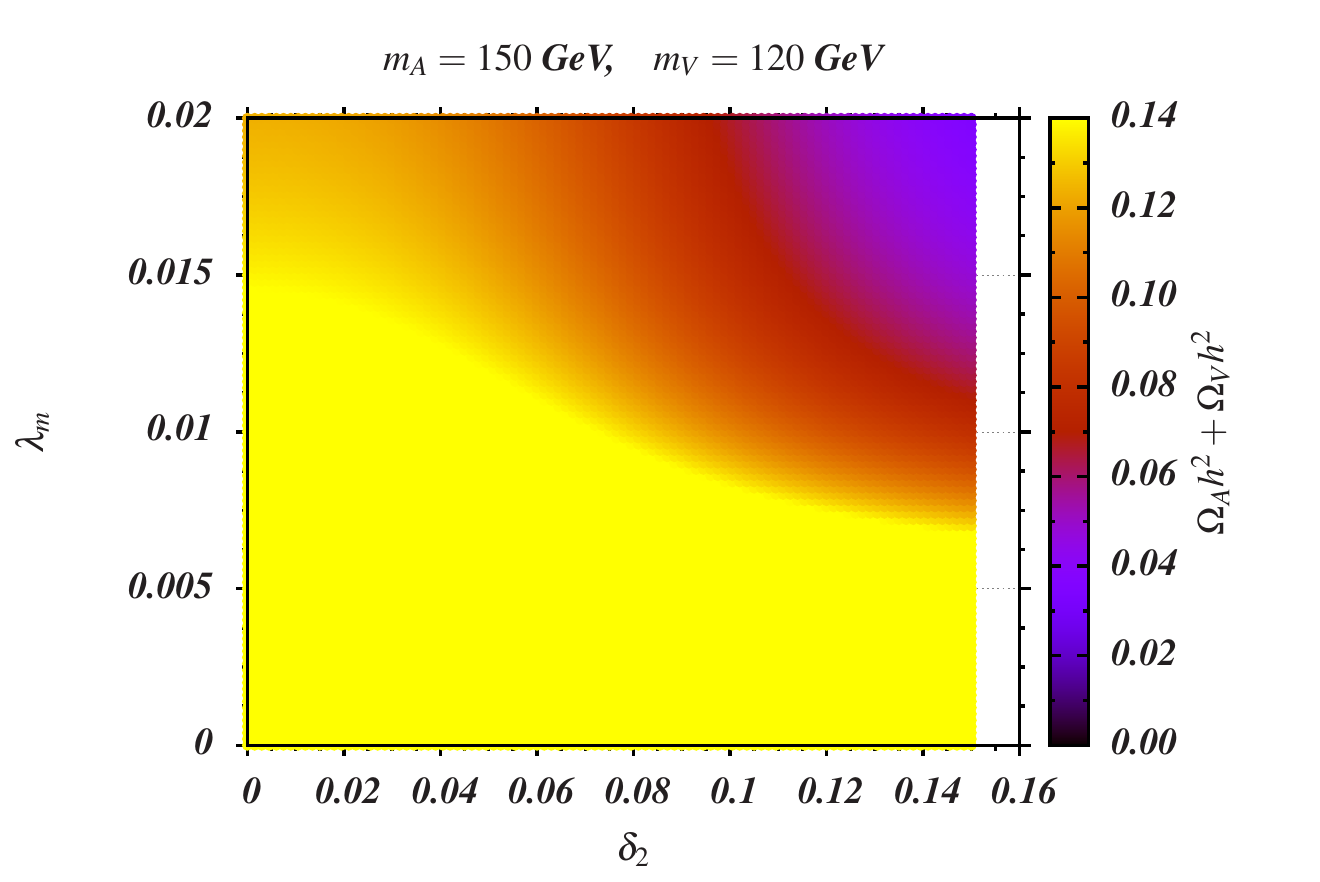} 
\end{minipage}
    \caption{Plots of $\delta_{2}$ and $\lambda_{m}$ $vs.$ relic density.}
    \label{fig:dlt2lmdm}
\end{figure}

\item{{\bf The third group: varying $\lambda_m$ and $\delta_2$ }}

FIG.~\ref{fig:dlt2lmdmcorsec} depicts the relationships among
$\delta_{2}$, $\lambda_{m}$ and the cross sections.
FIG.~\ref{fig:dlt2lmdm} illustrates that the magnitude of $\Omega_Ah^2$($\Omega_Vh^2$) reaches its 
critical value around $\delta_{2}\sim0.09$~($\lambda_{m}\sim0.01$). 
And that the DM relic density for each component almost depends on their own coupling parameters,
this is caused by the fact that all related annihilation cross sections are proportional to $\delta_{2}$ or $\lambda_{m}$.

\end{enumerate}

\subsection{Direct detection}
In our model,  the two component DMs, i.e., the scalar $A$ and vector $V$, interact with the SM particles 
through the exchange of three Higgs bosons. Thus, the DM-nucleon scattering cross section is spin-independent (SI). 
For each single component, the SI DM-nucleon cross sections are calculated to be
\begin{eqnarray}
&&\sigma_{SI}^{A}=\frac{1}{16\pi v^{2}}
\frac{m_{N}^{4}f_{N}^{2}}{(m_{A}+m_{N})^{2}}
\left|\frac{\delta_{2}}{2}v\frac{R_{11}}{m_{h_{1}}^{2}}
+\frac{\delta_{1}}{2}v_{\phi}\frac{R_{21}}{m_{h_{2}}^{2}}
+\frac{d_{2}}{2}v_{s}\frac{R_{31}}{m_{h_{3}}^{2}}\right|^{2},\\
&&\sigma_{SI}^{V}=\frac{1}{16\pi v^{2}}
\frac{m_{N}^{4}f_{N}^{2}}{(m_{V}+m_{N})^{2}}
\left|g_{\phi}^{2}v_{\phi}\frac{R_{21}R_{11}}{m_{h_{1}}^{2}}
+g_{\phi}^{2}v_{\phi}\frac{R_{22}R_{21}}{m_{h_{2}}^{2}}
+g_{\phi}^{2}v_{\phi}\frac{R_{23}R_{31}}{m_{h_{3}}^{2}}\right|^{2},
\end{eqnarray}
with $R_{ij}$ coming from Eq.~(\ref{Rvals}), $m_{N}$ and
$f_{N}=\sum f_{L}+3\times\frac{2}{27}f_{H}$ are the nucleon mass and effective Higgs-nucleon coupling respectively, 
where $f_{N}$ is the summation of light quark ($f_{L}$) and heavy quark ($f_{H}$) contributions, and we take the value 
$f_{N}=0.326$~\cite{Young:2009zb} in our numerical analysis. Since 
the current experiments assume that the local DM density is provided by one single DM specie, 
the situation that both $A,~V$ components contribute to 
the local DM density making us unable to use the current experimental results directly. 
Assuming the contribution of each DM component to the local density is the same as their contribution 
to the relic density, the SI scattering cross section should be rescaled by a factor 
${\Omega_{A,V}h^{2}}/{\Omega_{DM}h^{2}}$. Thus, the corresponding upper limit on the SI DM-nucleon cross section 
of each single component is~\cite{Belanger:2011ww, Barger:2008jx}:
\begin{eqnarray}
\sigma^{A}_{SI}\leq ({\Omega_{DM}h^{2}}/{\Omega_{S}h^{2}})\sigma^{exp}_{SI}(M_A)\, ,\\ 
\sigma^{V}_{SI}\leq ({\Omega_{DM}h^{2}}/{\Omega_{V}h^{2}})\sigma^{exp}_{SI}(M_V)\, .
\end{eqnarray}

Similar to the above Section, here we do analyses using the LUX experiment results~\cite{Akerib:2013tjd} via three groups as well.
\begin{figure}[!htb]
\begin{minipage}[t]{0.4\textwidth} 
    \centering
    \includegraphics[scale=0.6]{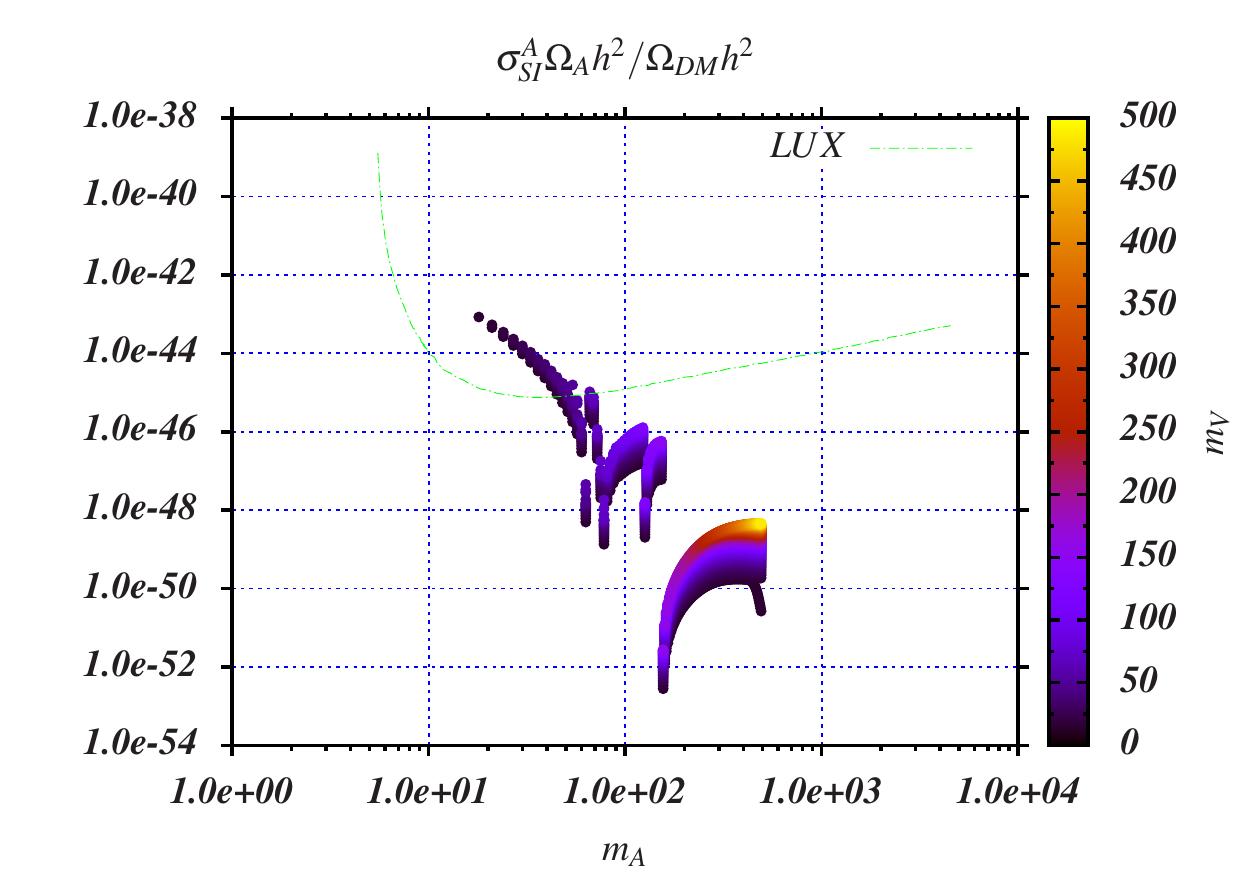} 
\end{minipage}
\hspace{0.1\textwidth}
\begin{minipage}[t]{0.4\textwidth} 
    \centering
    \includegraphics[scale=0.6]{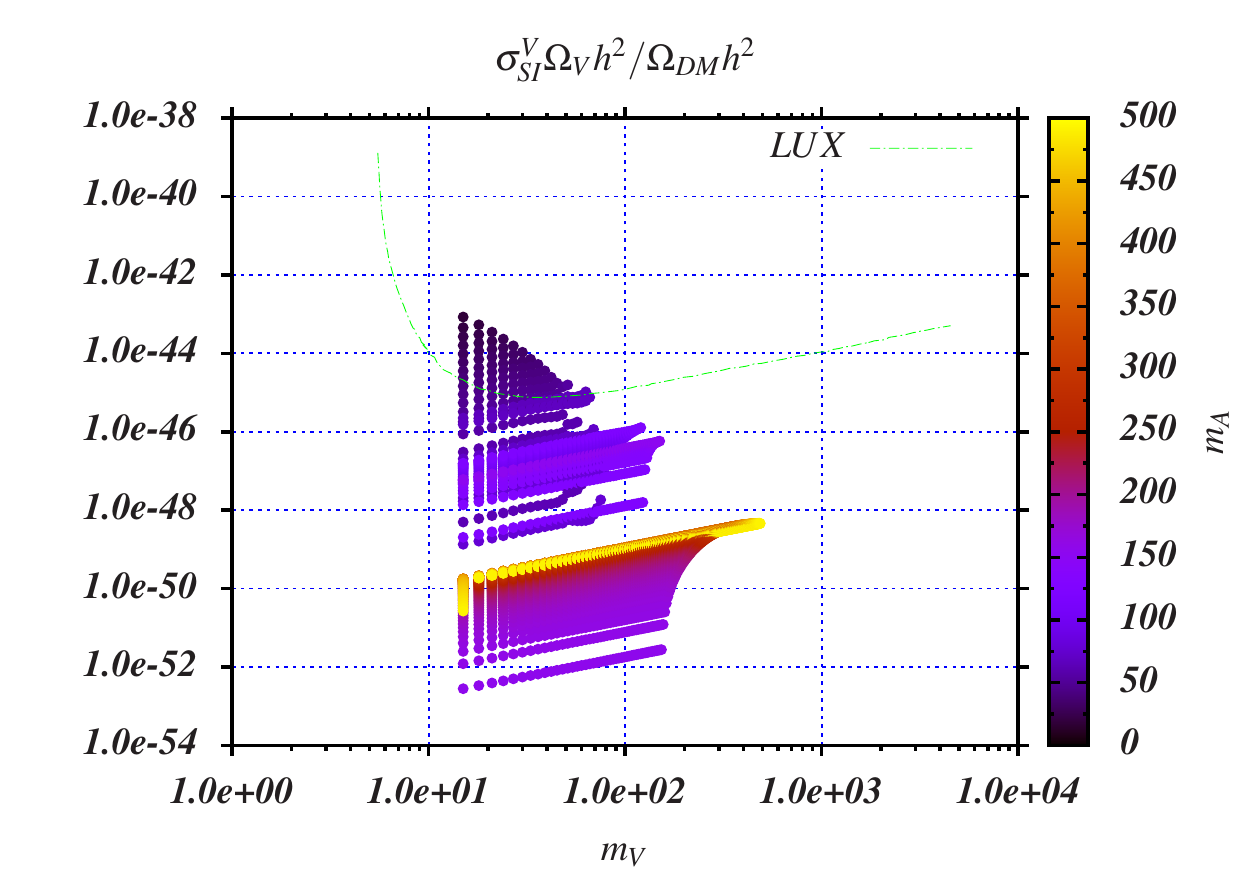} 
\end{minipage}
    \caption{For the case of $m_A>m_V$, parameter regions being left considering the direct detection constraint of LUX.}
    \label{fig:directdetmAmV}
\end{figure}
~\\
\begin{enumerate} 
\item{\bf $m_A>m_V$} 

The left and right panels of Fig.~\ref{fig:directdetmAmV}
depict that: with the value of $m_V$ being unbounded, $m_A$ should be bigger than $30$~GeV. 
\item{\bf $m_V>m_A$} 

The left and right panels of Fig.~\ref{fig:directdetmVmA}
depict that $m_{A,V}$ should be no smaller than $50$~GeV.

\begin{figure}[!htb]
\begin{minipage}[t]{0.4\textwidth} 
    \centering
    \includegraphics[scale=0.6]{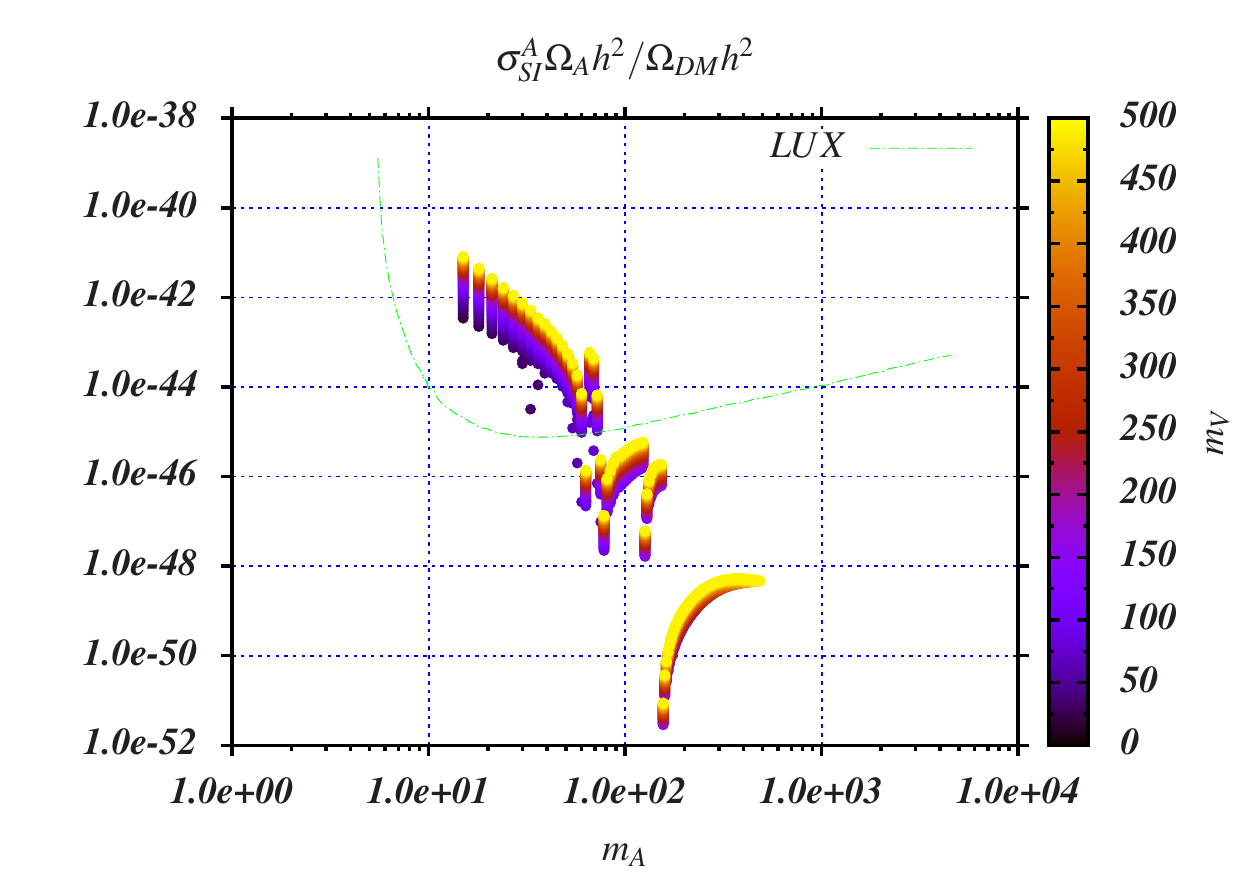} 
\end{minipage}
\hspace{0.1\textwidth}
\begin{minipage}[t]{0.4\textwidth} 
    \centering
    \includegraphics[scale=0.6]{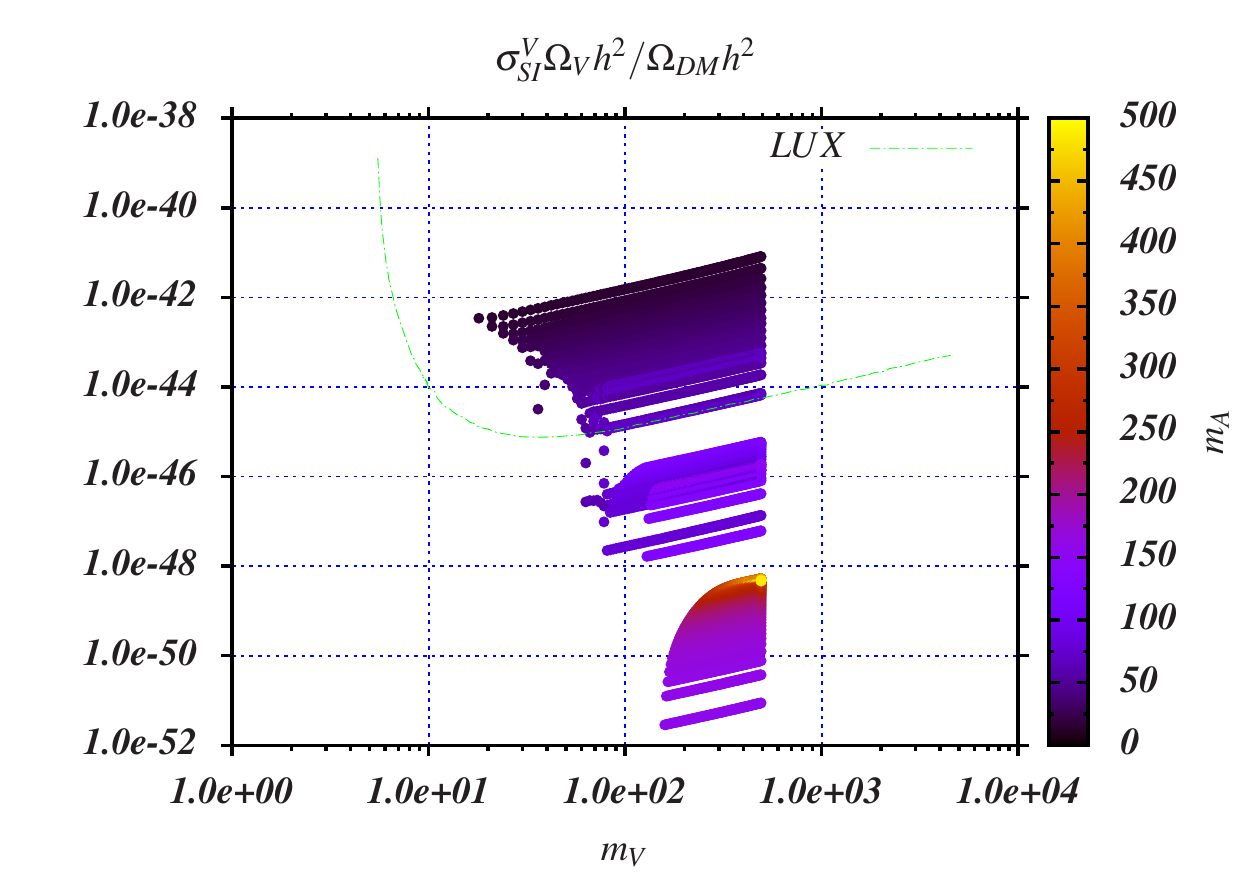} 
\end{minipage}
    \caption{For the case of $m_V>m_A$, parameter regions being left considering the direct detection constraint of LUX.}
    \label{fig:directdetmVmA}
\end{figure}
\begin{figure}[!h]
\begin{minipage}[t]{0.4\textwidth} 
    \centering
    \includegraphics[scale=0.6]{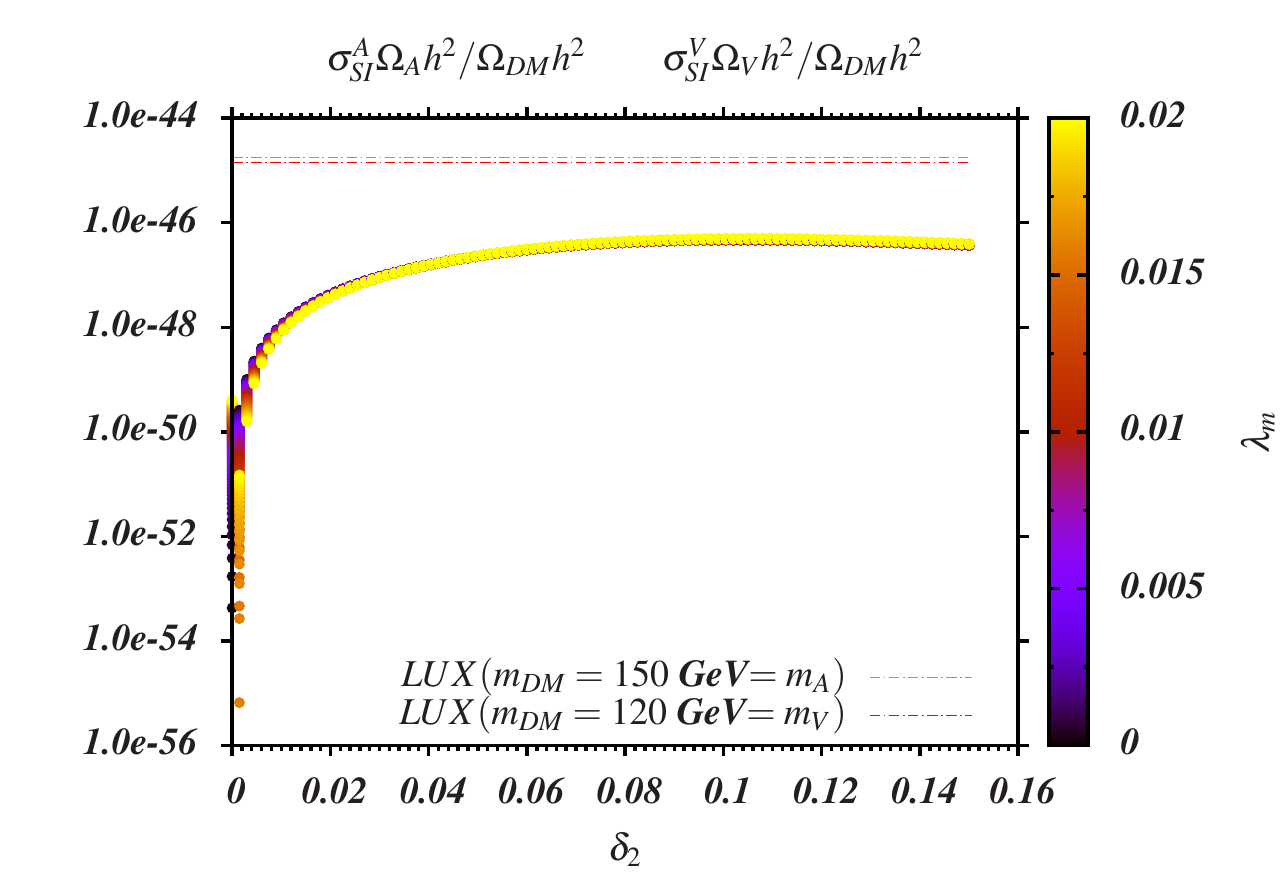} 
\end{minipage}
\hspace{0.1\textwidth}
\begin{minipage}[t]{0.4\textwidth} 
    \centering
    \includegraphics[scale=0.6]{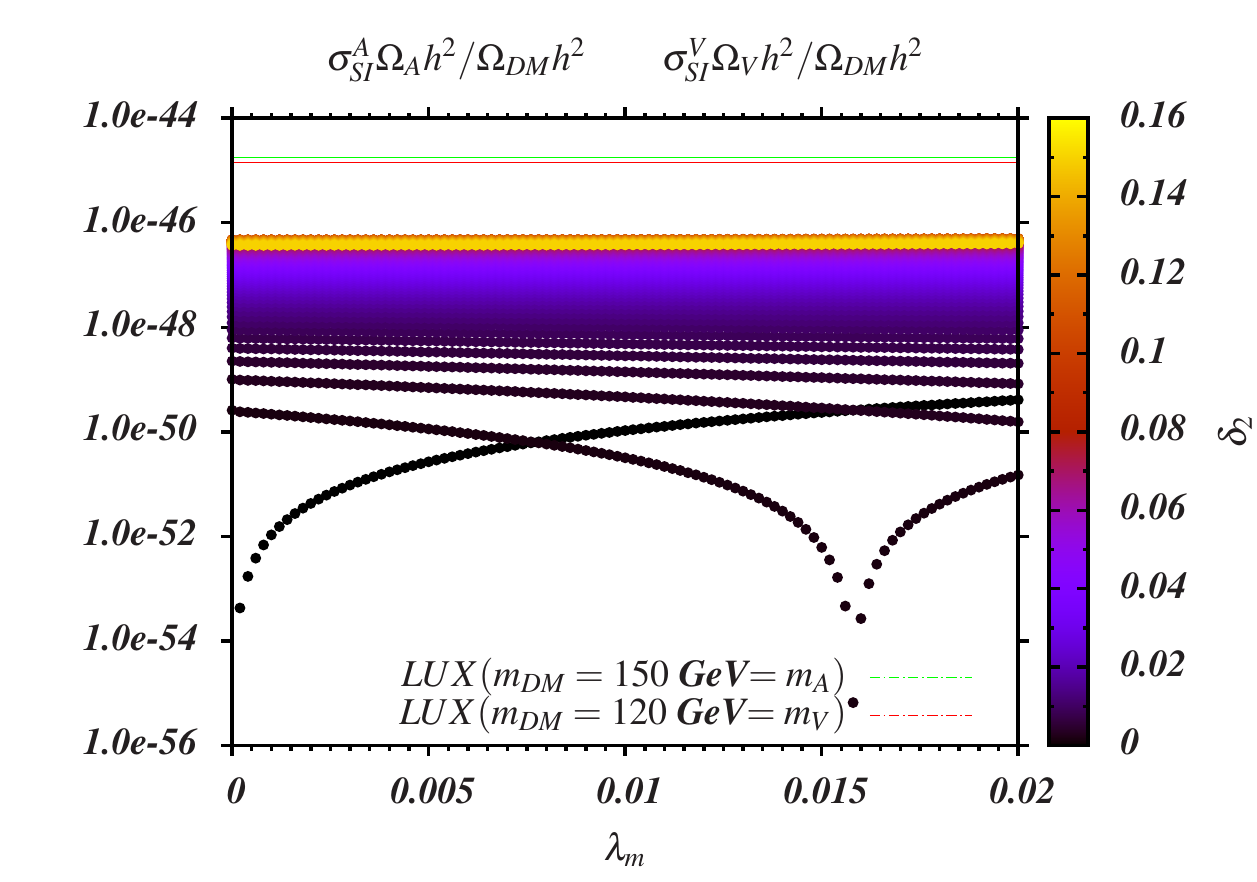} 
\end{minipage}
    \caption{$\delta_2$ and $\lambda_m$. Parameter regions being left considering the direct detection constraints on the model coefficients.}
    \label{fig:directdetcoeff}
\end{figure}

\item{ parameter space of $\delta_2$ and $\lambda_m$}

Fig.~\ref{fig:directdetcoeff} depicts that all the parameter spaces  
satisfy the experimental constraints. 

\end{enumerate}

\section{Higgs indirect search and electroweak precision constraints}

In this section, we study the effects of the mixing among the three fields $h,~\eta',~S$ using
the Higgs indirect search and constraints from electroweak precision observables.

\subsection{Higgs indirect search}

\begin{figure}[!h]
\begin{minipage}[t]{0.4\textwidth} 
    \centering
    \includegraphics[scale=0.6]{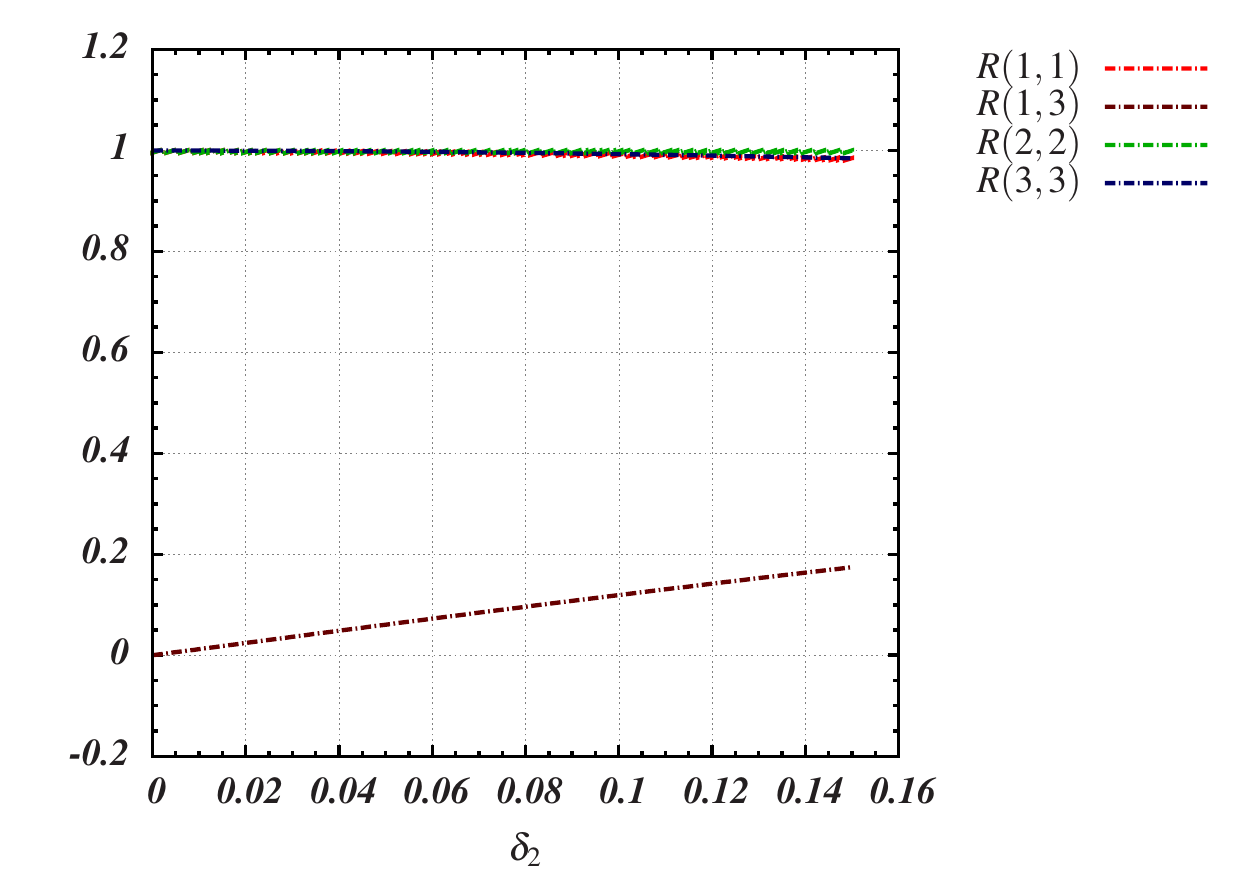} 
\end{minipage}
\hspace{0.1\textwidth}
\begin{minipage}[t]{0.4\textwidth} 
    \centering
    \includegraphics[scale=0.6]{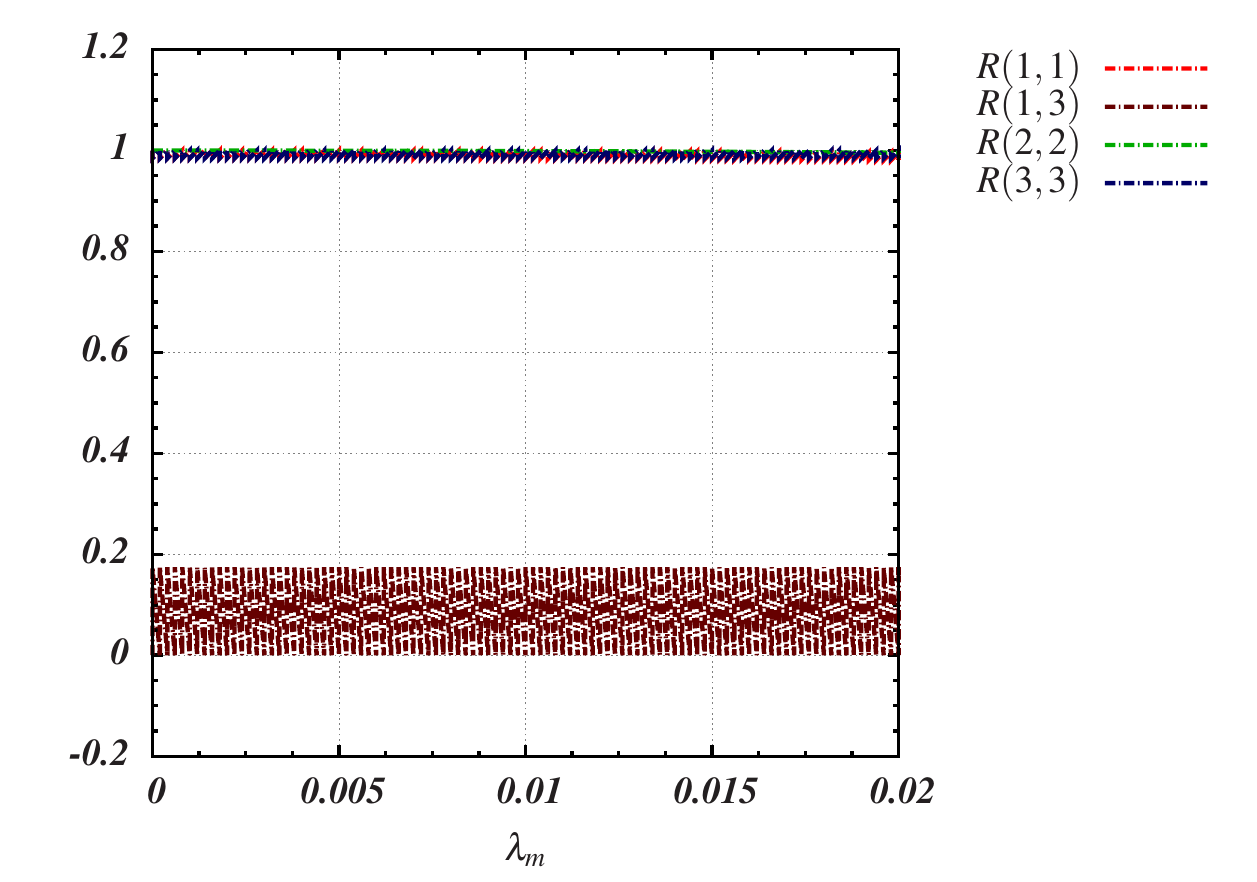} 
\end{minipage}
    \caption{$R$ values with respect to $\lambda_m$ and $\delta_2$.}
    \label{fig:R}
\end{figure}

The current status of the LHC measurements of the Higgs couplings constrains 
the matrix element $R_{11}$, which describes the discrepancy between
$h_1$ and the SM Higgs. The couplings of $h_1$ to all
SM particles are the rescaled values of the SM couplings, taking the
form of
\begin{eqnarray}
g_{h_1 XX}=R_{11}g_{h_1 XX}^{SM}\; .
\end{eqnarray}
For the first two group parameter spaces in the last Section, additional decay channel $h_1\rightarrow AA(VV)$
exists, which almost does not change the total width, since we have the square of $R_{12}(R_{13})$
to suppress the magnitude of $\Gamma(h_1\rightarrow AA(VV))$. The same logic applies to the analysis with the third parameter group
used in the last Section~\footnote{Because the mass of $h_2(h_3)$ is above the threshold, additional channel $h_1\rightarrow h_2h_2(h_3 h_3)$, 
which changes the total width, does
not open. For the study of one Higgs decays to two other Higgs bosons, we refer to~\cite{Kang:2013rj}. }.
Thus signal rates $\mu_{XX}$ associated with Higgs measurements are functions of
$R_{11}$
\begin{eqnarray}
\mu_{XX}=\frac{\sigma\cdot BR}{\sigma^{SM}\cdot BR^{SM}}=R_{11}^2\; ,
\end{eqnarray}
with $\sigma,~BR$ (that with a a superscript $SM$) being the production cross section and branching ratios of $h_1$ (the SM Higgs).
Therefore, to what extent the model differs from the SM Higgs measurements is determined by the value of $R_{11}$. 
The value of $R_{11}$ for the parameter setup given in Table.~\ref{tab:mAmV} is $0.9973$, 
and we refer to Fig.~\ref{fig:R} for the $R$ vales\footnote{Here, we use $R(i,j)$ as the labels in the two plots, which means the same as that of $R_{ij}$ in other place of this work.} corresponding to the benchmark scenario given in Table.~\ref{tab:VEV}.

\subsection{Electroweak precision observables constraints}

The presence of the scalar $h_3$ with mass $m_{h_3}<114$ GeV is subject to tight bounds from the LEP~\cite{Barate:2003sz},
while these bounds are highly released in all our parameter choices, since the mixing between $h_1$ and  $h_3$ is very small,
about $0.0573$ for benchmark scenario given in Table.~\ref{tab:mAmV} and 
could be seen in the Fig.~\ref{fig:R} for the benchmark scenario given in Table.~\ref{tab:VEV}.

As for electroweak precision observables, the oblique parameters $S$ and $T$ 
which parameterize potential new physics contributions 
to electroweak radiative corrections, are computed following Refs.~\cite{Chivukula:2013xka,Hambye:2008bq}. 
In our model, they are 
\begin{eqnarray}
\label{eq:ST}
S &=& \frac{1}{24\pi}\Biggl\{ R_{11}^{2}\left[\log R_{h_{1}h}
+G(m_{h_{1}}^{2},m_{Z}^{2})-G(m_{h_{1}}^{2},m_{Z}^{2})\right]\nonumber\\
&&\qquad +R_{12}^{2}\left[\log R_{h_{2}h}
+G(m_{h_{2}}^{2},m_{Z}^{2})-G(m_{h_{2}}^{2},m_{Z}^{2})\right]\nonumber\\
&&\qquad +R_{13}^{2}\left[\log R_{h_{3}h}
+G(m_{h_{3}}^{2},m_{Z}^{2})-G(m_{h_{3}}^{2},m_{Z}^{2})\right]\Biggr\},\\
T &=& \frac{3}{16\pi\sin^{2}\theta_{W}}\Biggl\{ R_{11}^{2}\left[\frac{1}{\cos^{2}\theta_{W}}
\left(\frac{\log R_{Zh_{1}}}{1-R_{Zh_{1}}}
-\frac{\log R_{Zh}}{1-R_{Zh}}\right)
-\left(\frac{\log R_{Wh_{1}}}{1-R_{Wh_{1}}}
-\frac{\log R_{Wh}}{1-R_{Wh}}\right)\right]\nonumber\\
&&\quad\quad+R_{12}^{2}\left[\frac{1}{\cos^{2}\theta_{W}}
\left(\frac{\log R_{Zh_{2}}}{1-R_{Zh_{2}}}
-\frac{\log R_{Zh}}{1-R_{Zh}}\right)
-\left(\frac{\log R_{Wh_{2}}}{1-R_{Wh_{2}}}
-\frac{\log R_{Wh}}{1-R_{Wh}}\right)\right]\nonumber\\
&&\quad\quad +R_{13}^{2}\left[\frac{1}{\cos^{2}\theta_{W}}
\left(\frac{\log R_{Zh_{3}}}{1-R_{Zh_{3}}}
-\frac{\log R_{Zh}}{1-R_{Zh}}\right)
-\left(\frac{\log R_{Wh_{3}}}{1-R_{Wh_{3}}}
-\frac{\log R_{Wh}}{1-R_{Wh}}\right)\right]\Biggr\},
\end{eqnarray}
where, $R_{AB}$, $G(m_{A}^{2},m_{B}^{2})$ and $f(R_{AB})$ are given by
\begin{eqnarray}
R_{AB} &=& \frac{m_{A}^{2}}{m_{B}^{2}}\; ,
\end{eqnarray}
\begin{eqnarray}
G(m_{A}^{2},m_{B}^{2}) &=& -\frac{79}{3}+9R_{AB}-2R_{AB}^{2}
+(12-4R_{AB}+R_{AB}^{2})f(R_{AB})\nonumber\\
&&+(-10+18R_{AB}-6R_{AB}^{2}+R_{AB}^{3}-9\frac{R_{AB}+1}{R_{AB}-1})\log R_{AB},\\
f(R_{AB}) &=& 
\begin{cases}
\begin{array}{cc}
\sqrt{R_{AB}(R_{AB}-4)}\log\left|\frac{R_{AB}-2-\sqrt{R_{AB}(R_{AB}-4)}}{2}\right| 
& R_{AB}>4,\\
0 & R_{AB}=4,\\
2\sqrt{R_{AB}(4-R_{AB})}\arctan\sqrt{\frac{4-R_{AB}}{R_{AB}}} & R_{AB}<4.
\end{array}
\end{cases}
\end{eqnarray}

\begin{figure}[!h]
%\begin{minipage}[t]{0.45\textwidth} 
%    \centering
    \includegraphics[width=0.9\textwidth]{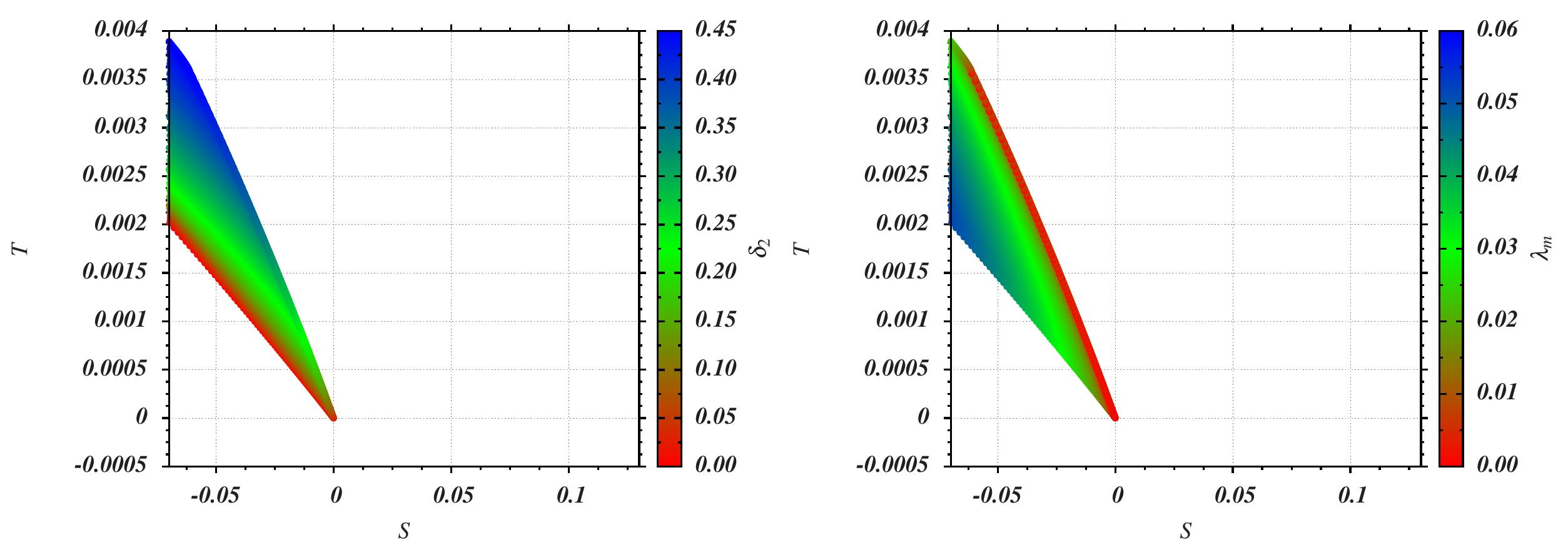}
%\end{minipage}
%\hspace{0.5cm}
    \caption{Constraints on $\delta_{2},\,\lambda_{m}$ from limits of S and T.}
    \label{fig:SToncoeff2}
\end{figure}
The parameters set is given in Table~\ref{tab:VEV}
and the magnitude of $g_{\phi}$ is determined by $m_V$ and $v_{\phi}$. 
The constraints on two parameters $\delta_{2}$ and $\lambda_{m}$ from $S$ and $T$ parameters are
shown in Fig.~\ref{fig:SToncoeff2}.
The $S$ and $T$ constraints require that $\delta_{2}$ and $\lambda_{m}$ should be smaller than about $0.45$ and $0.06$ respectively.
$S$ and $T$ are very sensitive to the mixing effects.
The first two parameter groups (as given in Section~\ref{sec:rda}) does not change the mixing angles among $h,\eta',
S$, and $S$ and $T$ give rise to the null limits since the mixing effects among $h$ and $\eta',~S$ in that cases are very small.

\section{Vacuum stability}
\label{VS}
The global minimum of the tree-level potential of our model requires,
\begin{equation}
\lambda >0\; , \lambda_\phi > 0\; , d_2 >0\; ,\lambda d_2 >4 \delta_2^2\; ,\lambda_\phi d_2 >\delta_1^2\; ,\lambda\lambda_\phi>16 \lambda_m^2\; .
\end{equation}
From one-loop renormalization group equations (RGEs) of the Higgs quartic couplings in Appendix C, 
i.e., Eq.~(\ref{beta}),
we find that the Higgs portals couplings $\lambda_m$ and $\delta_2$ are all get involved and give rise to positive contributions
to $\beta_\lambda$.
Thus, we could expect the vacuum stability problem being solved or alleviated in some parameter spaces.
Adopting the central values of top quark mass, the Higgs mass, and the strong coupling~\cite{PDG:relicdensity} 
as the low energy boundary conditions, we find that with the increasing energy scale, 
the Higgs quartic coupling running to a negative value around the scale
$10^9$ GeV, and then grows to positive values latter~\cite{Khosravi:2013lea,Degrassi:2012ry}. 

\begin{figure}[!htb]
\begin{minipage}[t]{0.5\textwidth} 
    \centering
    \includegraphics[scale=0.55]{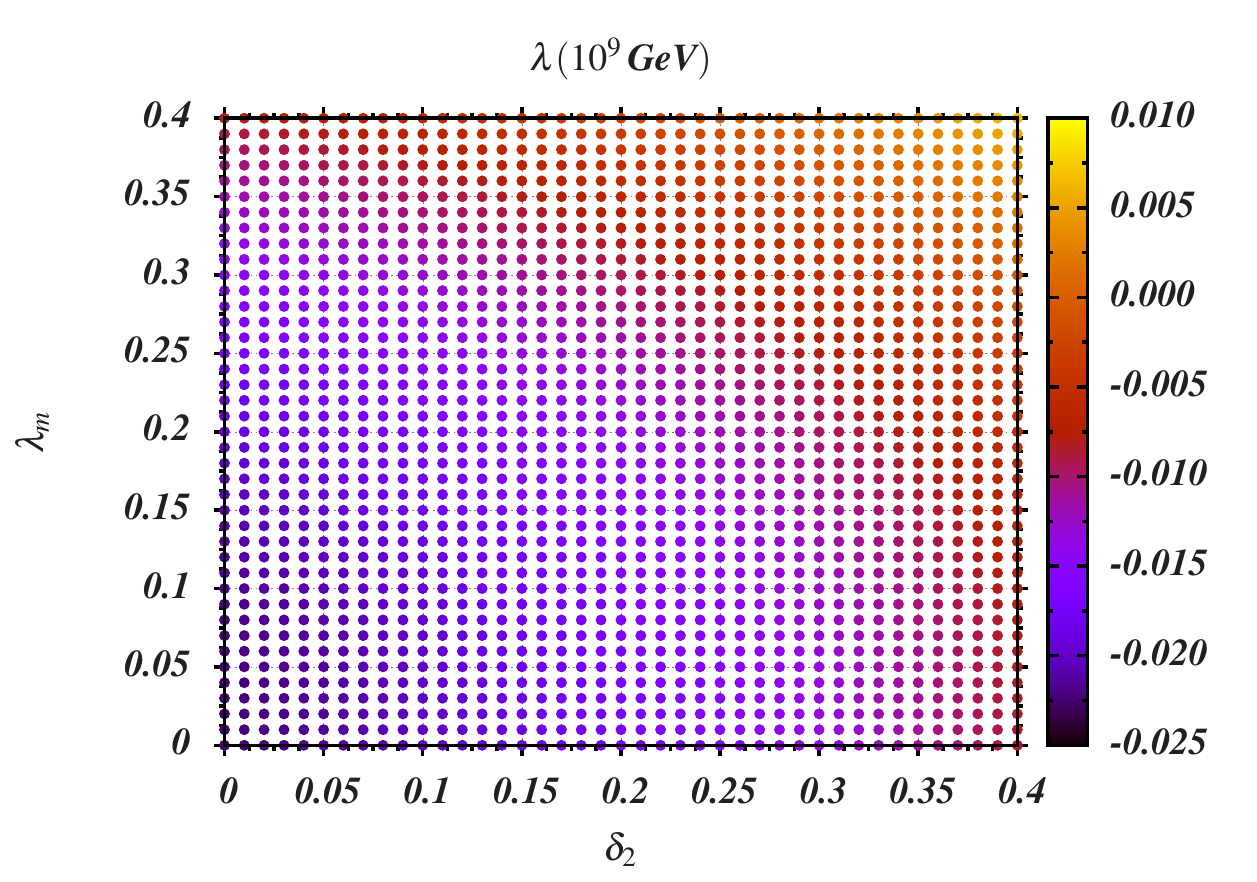}
\end{minipage}
\hspace{0.003\textwidth}
\begin{minipage}[t]{0.5\textwidth} 
    \centering
    \includegraphics[scale=0.55]{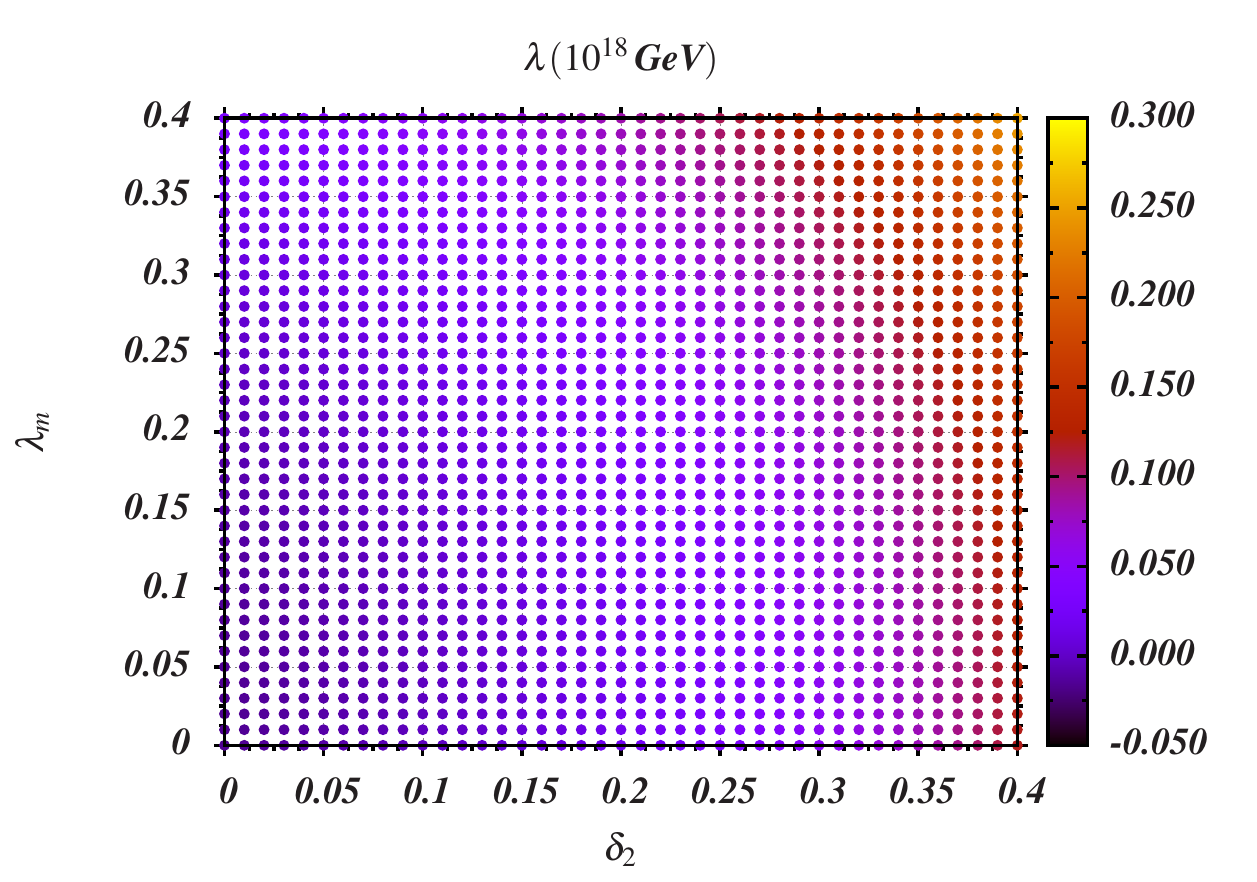}
\end{minipage}
    \caption{The Higgs quartic coupling $\lambda$ as a function of 
    $\lambda_m$ and $\delta_2$ at $10^9$ GeV (left panel) and $10^{18}$ GeV (right panel). With parameters set as: $\lambda=0.129$, $\lambda_{\phi}=0.2$, $d_2=0.2$, $\delta_1=0.2$, 
             $\delta_2=0.2$, $\lambda_{m}=0.2$\;.}
    \label{RR}
\end{figure}

Based on arguments on vacuum stability given in Ref.~\cite{Degrassi:2012ry}, 
one needs have positive value of the Higgs quartic coupling to ensure  absolute stability. 
It is obvious that to obtain absolute stability, we should elevate the curves of the Higgs quartic coupling
in the plot of $\lambda-\mu$~\cite{Khosravi:2013lea,Degrassi:2012ry}. 
From the $\beta$ functions given in Eq.~(\ref{beta}), we find that with the increasing of $\lambda_m$ and $\delta_2$ 
one could have the increasing of the value of the Higgs quartic coupling $\lambda$.
To verify this, we explore parameter spaces that survive under $S$ and $T$ limits as shown in Fig.~\ref{RR}, 
and we find that $\lambda_m$ and $\delta_2$ are required to be bigger than around $0.35$ in order to evade the stability problem.

\begin{figure}[!htb]
\label{fig:ns1}
\begin{minipage}[t]{0.5\textwidth} 
    \centering
    \includegraphics[scale=0.55]{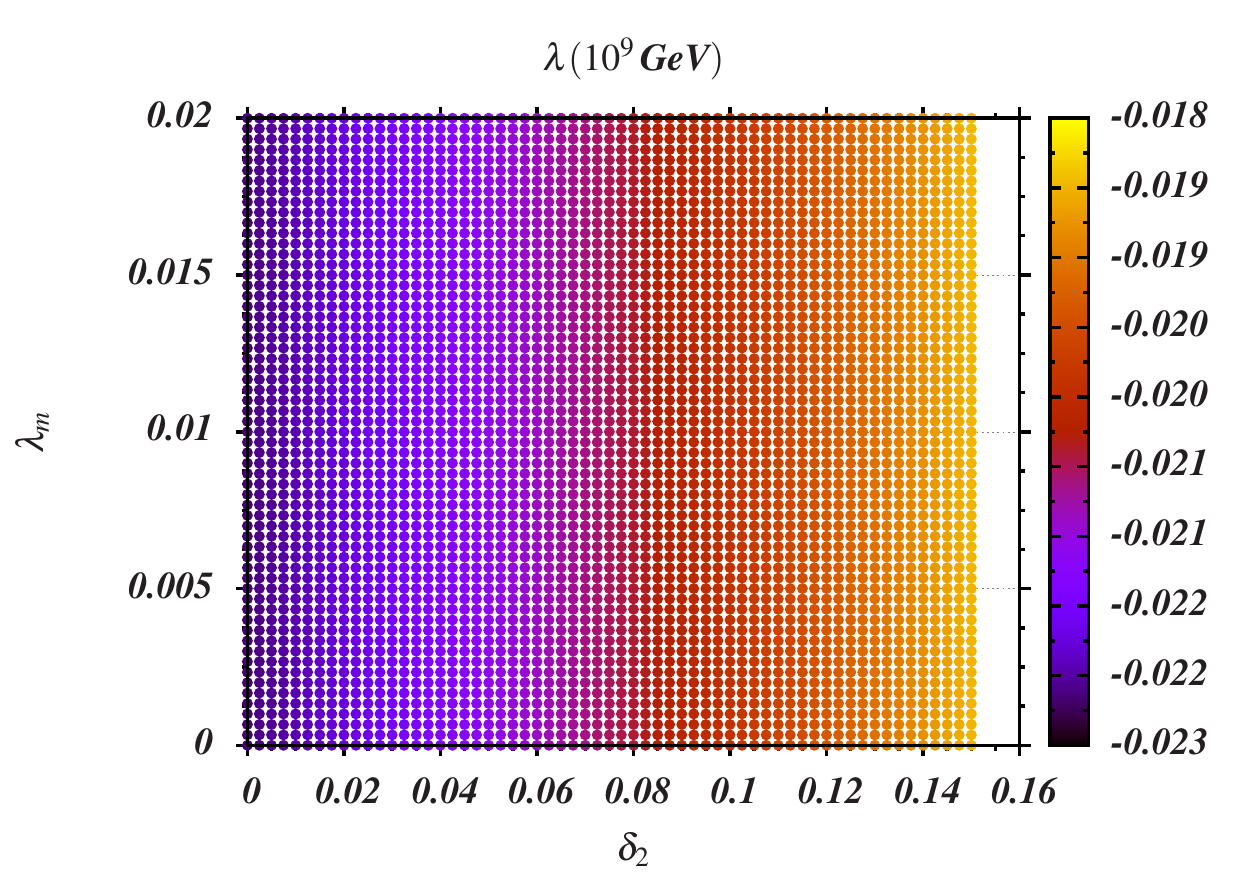}
\end{minipage}
\hspace{0.003\textwidth}
\begin{minipage}[t]{0.5\textwidth} 
    \centering
    \includegraphics[scale=0.55]{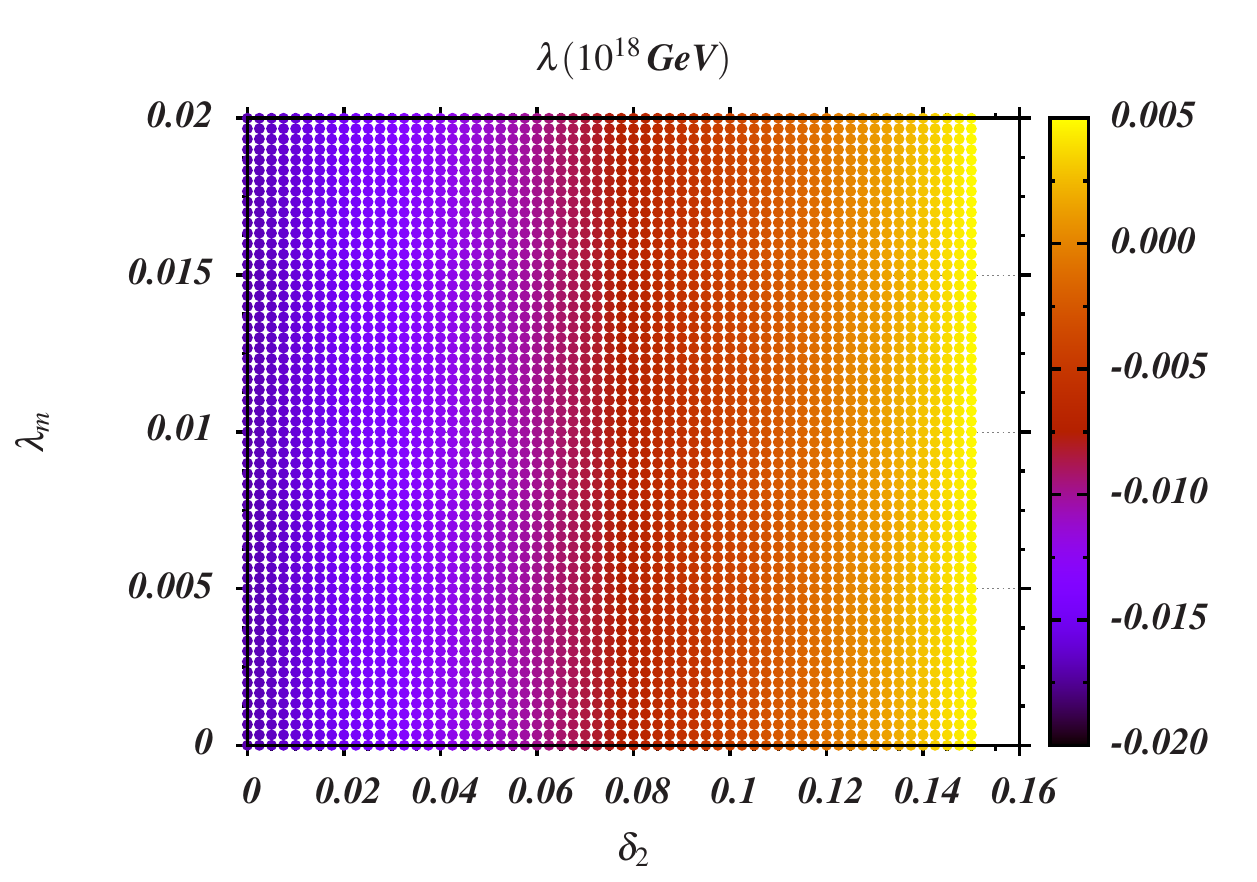}
\end{minipage}
    \caption{The Higgs quartic coupling $\lambda$ as a function of 
    $\lambda_m$ and $\delta_2$ at $10^9$ GeV (left panel) and $10^{18}$ GeV (right panel). With parameters set as: $\lambda=0.515$, $\lambda_{\phi}=0.2$, $d_2=0.2$, $\delta_1=0.02$, 
             $\delta_2=0.04$, $\lambda_{m}=0.01$\;.}
    \label{RR1}
\end{figure}

\begin{figure}[!htb]
%\begin{minipage}[t]{0.3\textwidth} 
    \centering
    \includegraphics[scale=0.6]{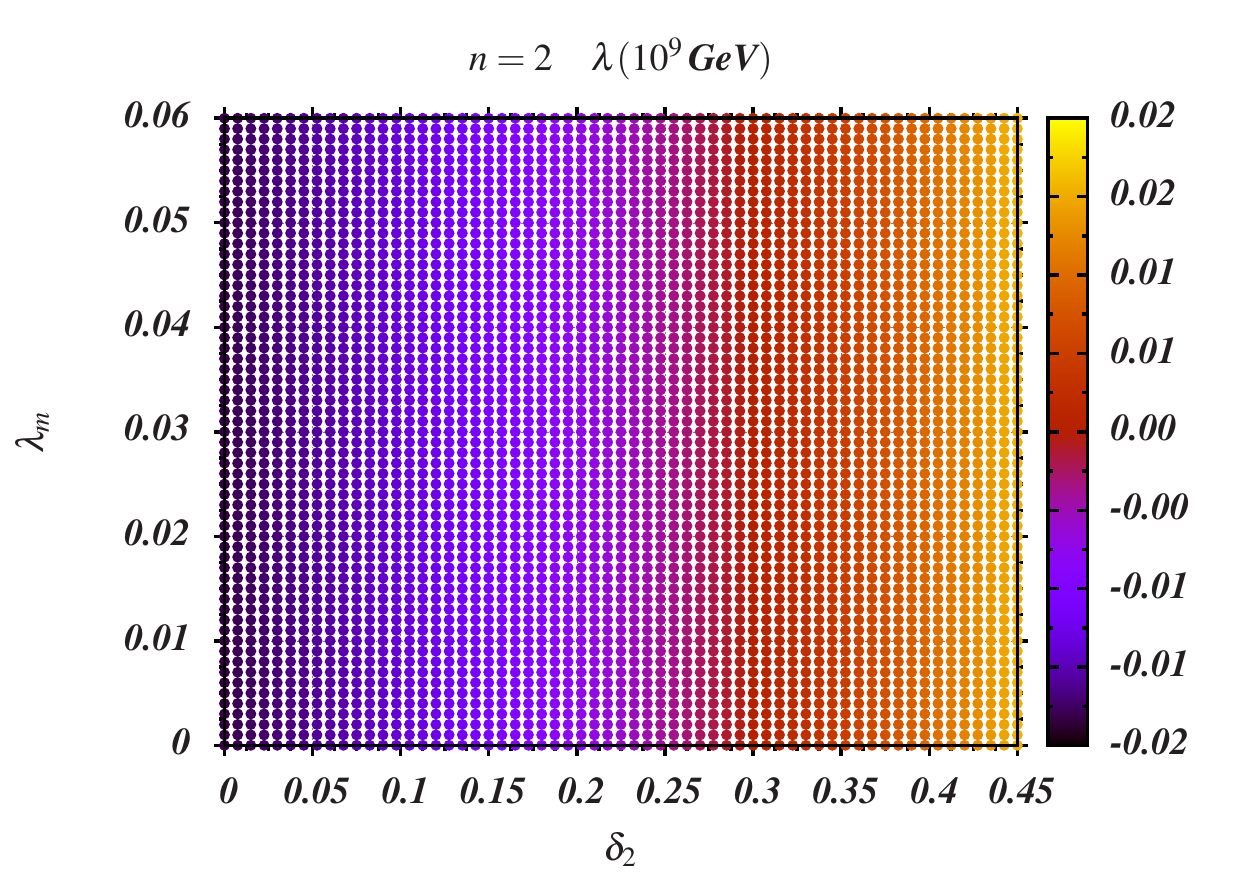} 
%\end{minipage}
 %\begin{minipage}[t]{0.3\textwidth} 
    \caption{$\lambda(10^{9}\,\mbox{GeV})$ as a function of $\delta_2$ and $\lambda_m$. $n=2$ is to indicate
    we have supplemented two $\bigS$.}
    \label{fig:stability}
\end{figure}

To figure out to what extent we 
need to lift the value of the Higgs quartic coupling in the third parameters setup (as given in Table~\ref{tab:VEV}), 
we plot contours of $\lambda(\mu=10^9$ ($10^{18}$) GeV) with respect to $\lambda_m$ and $\delta_2$, 
as shown in Fig.~\ref{RR1}, and we find that the vacuum is not bound from bellow there. 
The plot of $\lambda$ (at the scale of $10^9$ GeV) as a function of $\lambda_m$ and $\delta_2$, for the scenario with two $\bigS$ being supplemented (both two $\bigS$ have the same interaction
with the SM and the $SU(2)_D$ group), shows that with $\delta_2>0.25$ one can indeed obtain the potential bounded from below, as shown in Fig.~\ref{fig:stability}.
The two figures, i.e., Fig.~\ref{RR1} and Fig.~\ref{fig:stability}, illustrate that to achieve the vacuum stability, no smaller than two $\bigS$ is expected in the third parameters setup.


\section{Footprint of the naturalness problem}

In this section, we would like to present the relevant naturalness problem in our model, and 
discuss the indirect search of the scenario which could improve the naturalness problem.  

\subsection{The naturalness problem}
In the SM, considering the gauge invariant property of the two point Higgs Green function~\cite{Fleischer:1980ub}, 
the naturalness problem could be defined as
 \begin{eqnarray}
 \label{VCSM}
 (m^{0}_{h})^{2}=m^{2}_{h}+
 \frac{\Lambda^{2}}{(4\pi)^{2}}VC_{SM}\;,
 \end{eqnarray}
 where
 \begin{eqnarray}
  \label{VCSM}
 VC_{SM}= 12\lambda-12 g_{t}+\frac{9}{2}g_2^{2}+\frac{3}{2}g_1^2\; ,
 \end{eqnarray}
and $m_{h}$ ($m^{0}_{h}$) is the renormalized (bare) Higgs mass, $\Lambda$ indicates 
the cut off scale where the new fields are required to cancel the quadratic divergences of the Higgs mass square.
 
Although our model is renormalizable, we assume it as a low energy effective theory of 
some more fundamental theory, which is UV completed
at the scale $\Lambda$. Thus, Eq.~(\ref{VCSM}) becomes
\begin{eqnarray}
\label{VCVA} 
(m^{0}_{h})^{2}=m^{2}_{h}+
 \frac{\Lambda^{2}}{(4\pi)^{2}}VC_{VA}\;,
\end{eqnarray}
with
\begin{eqnarray}
\label{VCVA1} 
VC_{VA}&\approx&12\lambda+\frac{9}{2}g_2^2+\frac{3}{2}g_1^2-12g_t^2\nonumber\\
&+&\frac{5}{2}\lambda_m+\frac{\lambda_m^2}{\lambda_\phi}+\frac{3g_\phi^2 \lambda_m}{2\lambda_\phi}+n(\frac{3}{8}\delta_2+\frac{\delta_2\lambda_m}{\lambda_\phi}\nonumber\\
&+&\frac{\delta_2^2 d_2}{8}+\frac{\delta_2\delta_1}{8d_2})\; ,
\end{eqnarray}
in which new fields get involved. Here, we suppose that not only one $\bigS$ embracing the same property as explored in the model (thus $n$ is introduced to denote the number of $\bigS$), and the symbol $\approx$ is to indicate
that Eq.~(\ref{VCVA}) could be hold in the case where $S,~\eta'$, and $H$ have negligible mixing.

In addition, we would like to mention that, the VEV and Higgs mass in the SM are gauge invariant considering the renormalization of the mass term in the SM Higgs potential~\cite{Sirlin:1985ux}. And~\cite{Sirlin:1985ux} use the different tadpole renormalization method in comparison with that of~\cite{Fleischer:1980ub} and leaves no trails of tadpole contributions in 
 $VC_{SM}$ and $VC_{VA}$, 
which cast the new formula of
 \begin{eqnarray}
  \label{VCVAA}
 VC'_{SM}&= &-(6\lambda-6 g_{t}+\frac{9}{4}g_2^{2}+\frac{3}{4}g_1^2)\; ~,~\nonumber\\
 VC'_{VA}&\approx&-(6\lambda+\frac{9}{4}g_2^2+\frac{3}{4}g_1^2-6g_t^2)
-\frac{1}{2}\lambda_m-\frac{n}{2}\delta_2~.~\,
 \end{eqnarray}

To soften the naturalness problem~\cite{Grzadkowski:2009mj,Chakraborty:2012rb}, the simplest solution
 is to suppose $VC_{VA}=0$ (or $VC'_{VA}=0$) so that the modified Veltman condition~\cite{Veltman:1980mj}
is realized. For $\Lambda=10$ TeV, $n\approx 1( 6)$ is needed to obtain $VC_{VA}=0$ ($VC'_{VA}=0$).

\subsection{Indirect search for the scenario which alleviates the naturalness problem}

Considering $m_{\eta',S,A} \gg v$ could been satisfied in some parameter spaces
of the model~\footnote{For simplicity, we consider the case in which the mixings among the three fields
$\eta',S,A$ could be neglected.}, one may integrate out the $\eta',S,A$ and express
their effects in terms of an effective Lagrangian below the scale
$Min(m_{\eta'},m_S,m_A)$, which involves only the SM fields with appropriate
higher-dimensional operators. At one-loop level, integrating out the
$\eta',S,A$ leads to shifts in the wave-function renormalization and
the potential of the Higgs doublet $H$, as well as operators with dimension
six and higher.  The
dimension-six operators in the effective Lagrangian cast the form of
\begin{equation} 
\label{eq:Leff}
  \mathcal{L}_{eff} = \mathcal{L}_{SM} +( \frac{c^{\eta'}_H}{m_{\eta'}^2}+\frac{c^s_H}{m_S^2}+\frac{c^A_H}{m_A^2}) 
  \left( \frac{1}{2} \partial_\mu |H|^2 \partial^\mu |H|^2 \right)\; .
\end{equation}
Matching to the full theory at
the scale $m_{\eta',s,A}$, we have 
\begin{eqnarray}
c^{\eta'}_H &=& \frac{n_{\eta'}}{4} |\lambda_m|^2/(96\pi^2) \, ,  \nonumber\\
c^S_H &=&\frac{ n}{8} |\delta_2|^2/(96\pi^2  )\, ,\nonumber\\
c^A_H &=&\frac{ n}{8} |\delta_2|^2/(96\pi^2)  \, .
\end{eqnarray}

Below the scale of electroweak symmetry breaking, \Eq{eq:Leff}
leads to a shift in the wave-function renormalization of the physical
scalar $h$~\cite{Craig:2013xia}, with $\delta Z_h = 2 v^2 (c^{\eta'}_H  /
m_{\eta'}^2+c^S_H  /
m_S^2+c^A_H  /
m_A^2)$. After canonically normalizing $h$, i.e., $h\rightarrow (1-\delta Z_h /2)h$, 
its coupling to vectors and fermions are altered, which may lead to a measurable correction to, e.g., the $hZ$
associated production cross-section
\begin{equation}
\label{apc}
  \delta \sigma_{Zh} = - 2 v^2 (c^{\eta'}_H  /
m_\eta'^2+c^S_H  /
m_S^2+c^A_H  /
m_A^2) \;,
\end{equation}
where we have defined $\delta \sigma_{Zh}$ as the fractional change in the associated production 
cross section relative to the SM prediction, which by design vanishes for the case of the SM. 
\footnote{Here, we would like to mention that, 
which one of $c^\eta_H,~c^s_H,~c^A_H$ in the our model takes part in $\delta \sigma_{Zh}$ depends
on parameter choices, i.e, the derivations of  $c^{\eta'}_H,~c^S_H,~c^A_H$ require $m_{\eta',S,A} \gg v$.} 
The sizable $n_{\eta'}(n)$, being required to relax the naturalness problem~\footnote{
The sizable $n$ is also necessary from the viewpoint of the vacuum stability, see Section~\ref{VS}. } (see Eq.~(\ref{VCVA})), 
may causes the correspondingly 
observable effect in the precision measurement of $\sigma_{Zh}$, which is what the future lepton colliders supposed to detect. 
At last, if one use our model to analyze electroweak phase transition, the measurement of associated production cross section
of $Zh$ ($\sigma_{Zh}$) might imposes very strong constraint on the multi-Higgs portals couplings $\lambda_m$ and $\delta_2$.  
 
\section{Discussions and conclusion}

In our model, the vector field ($V$) and the imaginary part of complex singlet ($A$) are
stable due to the reduced custodial symmetry $SO(3)$ and the residual $Z_2$ symmetry. 
The interactions between the two DMs and resonant effects mediated by 
three Higgs portals are demonstrated through the study of the relic density
behavior with respect to $m_A$ and $m_V$. The effects of the multi-Higgs portals couplings on
the generation of DM relic density are shown in the parameter space of $\lambda_m$ and $\delta_2$.
The parameter spaces 
of $\lambda_m$ and $\delta_2$ are totally free under  the LUX experiment results limits. 
Constraints (from LUX experimental results) on parameter spaces of $m_A-m_V$ give rise to $m_A> 30$ GeV 
 ($m_A(m_V)> 50$ GeV) for the case of $m_A>m_V$ ($m_A<m_V$), which allows more relaxed parameter spaces
in compare with the simple models. As for the third parameter group: 
the electroweak precision observables require $\lambda_m<0.06$ and $\delta_2<0.45$,
the vacuum stability at high scale requires at least two complex singlet, which is supposed to
be consistent with the naturalness problem argument. 
Other benchmark scenario with one $\bigS$, which survives under indirect Higgs search and electroweak 
precision test limits, and could improve   
the stability problem is also explored .

At last, we would like to mention that, 
through tuning of the parameter $\delta_2$ associate with
tuning of $v_s$ in the first setup of parameter spaces~\footnote{In order to maintain small mixing between $h$ and $S$ which helps to escape constraints of LHC and electroweak observable experiments. 
  },  one could use the model to explain Galactic Center Excess observed by the Fermi Telescope, with $\langle\sigma v\rangle\approx 10^{-26}~ cm^3/s$ being provided by scalar DM pairs annihilating to $b\bar{b}$ and the relic density supplied by the vector DM\footnote{For a viable annihilating multi-component dark
matter model consisting of two real gauge singlet scalars, one can explain the low energy ($1-3$~GeV) gamma ray excess from both Galactic Centre and Fermi Bubble~\cite{Modak:2013jya}.}.  Details of which, and the realization of SFOEWPT and inflation in the model are left for further studies.  

\appendix
\section{Annihilation Cross Sections}
\label{annh}
When the Higgs mass $m_{h_{i}}$ is larger than twice of the SM particle masses, 
the corresponding visible decay channels widths are
\begin{eqnarray}
&&\Gamma_{h_{i}\to ff}=\frac{N_{c}G_{F}m_{f}^{2}}{4\sqrt{2}\pi}m_{h_{i}}
\left|R_{i1}\right|^{2}\left(1-\frac{4m_{f}^{2}}{m_{h_{i}}^{2}}\right)^{3/2},\\
&&\Gamma_{h_{i}\to WW}=\frac{G_{F}}{8\sqrt{2}\pi}m_{h_{i}}^{3}
\left|R_{i1}\right|^{2}\sqrt{1-\frac{4m_{W}^{2}}{m_{h_{i}}^{2}}}
\left(1-\frac{4m_{W}^{2}}{m_{h_{i}}^{2}}+\frac{12m_{W}^{4}}{m_{h_{i}}^{4}}\right),\\
&&\Gamma_{h_{i}\to ZZ}=\frac{G_{F}}{16\sqrt{2}\pi}m_{h_{i}}^{3}
\left|R_{i1}\right|^{2}\sqrt{1-\frac{4m_{Z}^{2}}{m_{h_{i}}^{2}}}
\left(1-\frac{4m_{Z}^{2}}{m_{h_{i}}^{2}}+\frac{12m_{Z}^{4}}{m_{h_{i}}^{4}}\right)\;,
\end{eqnarray}
and when the Higgs mass $m_{h_{i}}$ is larger than the twice of the DM mass, 
the invisible decay channels widths are
\begin{eqnarray}
&&\Gamma_{h_{i}\to VV}=\Biggl|R_{i2}g_{\phi}^{2}\frac{v_{\phi}}{v}\Biggr|^{2}\frac{v^{2}m_{h_{i}}^{3}}{128\pi m_{V}^{4}}\sqrt{1-\frac{4m_{V}^{2}}{m_{h_{i}}^{2}}}\left(1-\frac{4m_{V}^{2}}{m_{h_{i}}^{2}}+\frac{12m_{V}^{4}}{m_{h_{i}}^{4}}\right),\\
&&\Gamma_{h_{i}\to AA}=\Biggl|R_{i1}\frac{\delta_{2}}{2}+R_{i2}\frac{\delta_{1}}{2}\frac{v_{\phi}}{v}+R_{i3}\frac{d_{2}}{2}\frac{v_{s}}{v}\Biggr|^{2}\frac{v^{2}}{32\pi m_{h_{i}}}\sqrt{1-\frac{4m_{A}^{2}}{m_{h_{i}}^{2}}}\;.
\end{eqnarray}
The total decay widthes of the Higgs with mass $m_{h_{i}}$ are 
the summations of visible and invisible decay widths.

The annihilation cross sections for DM corresponding to Fig.~\ref{fig:feyndia} are given as follows:

%\begin{enumerate}
\begin{itemize}
\item{{\bf Top-left}}
\begin{eqnarray}
<\sigma v>_{VV\to h_{j}h_{j}}^{s}&=&\frac{1}{32\pi}\frac{g_{\phi}^{4}v_{\phi}^2}{m_{V}^{2}}
\sqrt{1-\frac{m_{h_{1}}^{2}}{m_{V}^{2}}}\nonumber\\
&\times&\left|
\left[\sum_{i}\frac{R_{2i}\left(R_{i1}R_{1j}\lambda v
+R_{i2}R_{2j}\lambda_{m}v_{\phi}
+R_{i3}R_{3j}\frac{\delta_{2}}{2}v_{s}\right)}
{4m_{V}^{2}-m_{h_{i}}^{2}+im_{h_{i}}\Gamma_{h_{i}}}
\right]\right|^{2},\nonumber\\
<\sigma v>_{VV\to ff}&=&n_{c}\sum_{f}\frac{m_{f}^{2}}{16\pi}\left(1-\frac{m_{f}^{2}}{m_{V}^{2}}\right)^{3/2}\left|\sum_{i}
\frac{(g_{\phi}^{2}v_{\phi}/v)R_{2i}R_{i1}}{4m_{V}^{2}-m_{h_{i}}^{2}+im_{h_{i}}\Gamma_{h_{i}}}\right|^{2},\nonumber
\end{eqnarray}

\begin{eqnarray}
<\sigma v>_{VV\to ZZ}&=&\frac{m_{V}^{2}}{128\pi}\sqrt{1-\frac{m_{Z}^{2}}{m_{V}^{2}}}\left(1-\frac{m_{Z}^{2}}{m_{V}^{2}}+\frac{3m_{Z}^{4}}{m_{V}^{4}}\right)
\left|\sum_{i}\frac{(g_{\phi}^{2}v_{\phi}/v)R_{2i}R_{i1}}{4m_{V}^{2}-m_{h_{i}}^{2}+im_{h_{i}}\Gamma_{h_{i}}}\right|^{2},\nonumber\\
<\sigma v>_{VV\to WW}&=&\frac{m_{V}^{2}}{64\pi}\sqrt{1-\frac{m_{W}^{2}}{m_{V}^{2}}}\left(1-\frac{m_{W}^{2}}{m_{V}^{2}}+\frac{3m_{W}^{4}}{m_{V}^{4}}\right)\left|\sum_{i}\frac{(g_{\phi}^{2}v_{\phi}/v)R_{2i}R_{i1}}{4m_{V}^{2}-m_{h_{i}}^{2}+im_{h_{i}}\Gamma_{h_{i}}}\right|^{2},\nonumber\\
<\sigma v>_{AA\to h_{j}h_{j}}^{s}&=&\frac{1}{32\pi}
\frac{1}{m_{A}^{2}}\sqrt{1-\frac{m_{h_{1}}^{2}}{m_{A}^{2}}}
\nonumber\\
&\times&\Biggl|\frac{\delta_{2}}{2}v
\left[\sum_{i}\frac{R_{1i}\left(R_{i1}R_{1j}\lambda v
+R_{i2}R_{2j}\lambda_{m}v_{\phi}
+R_{i3}R_{3j}\frac{\delta_{2}}{2}v_{s}\right)}{4m_{A}^{2}
-m_{h_{i}}^{2}+im_{h_{i}}\Gamma_{h_{i}}}\right]\nonumber\\
&+&\frac{\delta_{1}}{2}v_{\phi}\left[\sum_{i}
\frac{R_{2i}\left(R_{i1}R_{1j}\lambda v
+R_{i2}R_{2j}\lambda_{m}v_{\phi}
+R_{i3}R_{3j}\frac{\delta_{2}}{2}v_{s}\right)}{4m_{A}^{2}
-m_{h_{i}}^{2}+im_{h_{i}}\Gamma_{h_{i}}}\right]\nonumber\\
&+&\frac{d_{2}}{2}v_{s}\left[\sum_{i}
\frac{R_{3i}\left(R_{i1}R_{1j}\lambda v
+R_{i2}R_{2j}\lambda_{m}v_{\phi}
+R_{i3}R_{3j}\frac{\delta_{2}}{2}v_{s}\right)}{4m_{A}^{2}
-m_{h_{i}}^{2}+im_{h_{i}}\Gamma_{h_{i}}}\right]
\Biggr|^{2}\; ,\nonumber\\
<\sigma v>_{AA\to ff}&=&n_{c}\sum_{f}\frac{m_{f}^{2}}{16\pi}\left(1-\frac{m_{f}^{2}}{m_{A}^{2}}\right)^{3/2}\Biggl|\left[\sum_{i}
\frac{(\delta_{2}/2)R_{1i}R_{i1}}
{4m_{A}^{2}-m_{h_{i}}^{2}+im_{h_{i}}\Gamma_{h_{i}}}\right]
\nonumber\\
&\times&\left[\sum_{i}
\frac{(\delta_{1}v_{\phi}/2v)R_{2i}R_{i1}}{4m_{A}^{2}-m_{h_{i}}^{2}+im_{h_{i}}\Gamma_{h_{i}}}\right]
+\left[\sum_{i}\frac{(d_{2}v_{s}/2v)R_{3i}R_{i1}}
{4m_{A}^{2}-m_{h_{i}}^{2}+im_{h_{i}}\Gamma_{h_{i}}}\right]\Biggr|^{2},\nonumber\\
<\sigma v>_{AA\to ZZ}&=&\frac{m_{A}^{2}}{128\pi}\sqrt{1-\frac{m_{Z}^{2}}{m_{A}^{2}}}
\left(1-\frac{m_{Z}^{2}}{m_{A}^{2}}+\frac{3m_{Z}^{4}}{m_{A}^{4}}\right)
\times\Biggl|\left[\sum_{i}\frac{(\delta_{2}/2)R_{1i}R_{i1}}
{4m_{A}^{2}-m_{h_{i}}^{2}+im_{h_{i}}\Gamma_{h_{i}}}\right]\nonumber\\
&+&\left[\sum_{i}
\frac{(\delta_{1}v_{\phi}/2v)R_{2i}R_{i1}}
{4m_{A}^{2}-m_{h_{i}}^{2}+im_{h_{i}}\Gamma_{h_{i}}}\right]
+\left[\sum_{i}\frac{(d_{2}v_s/2v)R_{3i}R_{i1}}
{4m_{A}^{2}-m_{h_{i}}^{2}+im_{h_{i}}\Gamma_{h_{i}}}\right]\Biggr|^{2},\nonumber\\
<\sigma v>_{AA\to WW}&=&\frac{m_{A}^{2}}{64\pi}\sqrt{1-\frac{m_{W}^{2}}{m_{A}^{2}}}
\left(1-\frac{m_{W}^{2}}{m_{A}^{2}}+\frac{3m_{W}^{4}}{m_{A}^{4}}\right)
\times\Biggl|\left[\sum_{i}\frac{(\delta_{2}/2)R_{1i}R_{i1}}
{4m_{A}^{2}-m_{h_{i}}^{2}+im_{h_{i}}\Gamma_{h_{i}}}\right]\nonumber\\
&+&\left[\sum_{i}
\frac{(\delta_{1}v_{\phi}/2v)R_{2i}R_{i1}}{4m_{A}^{2}-m_{h_{i}}^{2}+im_{h_{i}}\Gamma_{h_{i}}}\right]
+\left[\sum_{i}\frac{(d_{2}v_{s}/2v)R_{3i}R_{i1}}
{4m_{A}^{2}-m_{h_{i}}^{2}+im_{h_{i}}\Gamma_{h_{i}}}\right]\Biggr|^{2}\; .\nonumber
\end{eqnarray}

\item{{\bf Top-right}}
\begin{eqnarray}
\label{VVAA}
<\sigma v>_{VV\to AA}=\frac{1}{32\pi}\frac{1}{m_{V}^{2}}\sqrt{1-\frac{m_{A}^{2}}{m_{V}^{2}}}\left|\left[\sum_{i}\frac{(g_{\phi}^{2}v_{\phi})R_{2i}\left(R_{i1}\frac{\delta_{2}}{2}v+R_{i2}\frac{\delta_{1}}{2}v_{\phi}+R_{i3}\frac{d_{2}}{2}v_{s}\right)}{4m_{V}^{2}-m_{h_{i}}^{2}+im_{h_{i}}\Gamma_{h_{i}}}\right]\right|^{2},\nonumber
\end{eqnarray}

\begin{eqnarray}
\label{AAVV}
<\sigma v>_{AA\to VV}=\frac{m_{A}^{2}}{128\pi}
\sqrt{1-\frac{m_{V}^{2}}{m_{A}^{2}}}\left(1-\frac{m_{V}^{2}}{m_{A}^{2}}
+\frac{3m_{V}^{4}}{m_{A}^{4}}\right)
\times\Biggl|\left[\sum_{i}\frac{(\delta_{2}/2)R_{1i}R_{i1}}
{4m_{A}^{2}-m_{h_{i}}^{2}+im_{h_{i}}\Gamma_{h_{i}}}\right]\nonumber
\end{eqnarray}
\begin{eqnarray}
\qquad\qquad\qquad+\left[\sum_{i}
\frac{(\delta_{1}v_{\phi}/2v)R_{2i}R_{i1}}{4m_{A}^{2}-m_{h_{i}}^{2}+im_{h_{i}}\Gamma_{h_{i}}}\right]
+\left[\sum_{i}\frac{(d_{2}v_{s}/2v)R_{3i}R_{i1}}
{4m_{A}^{2}-m_{h_{i}}^{2}+im_{h_{i}}\Gamma_{h_{i}}}\right]\Biggr|^{2}.\nonumber
\end{eqnarray}
\item{{\bf Bottom-left}}
\begin{eqnarray}
<\sigma v>_{VV\to h_{i}h_{i}}^{t+u}&=&\frac{1}{4\pi}\frac{1}{m_{V}^{2}}
\sqrt{1-\frac{m_{h_i}^{2}}{m_{V}^{2}}}
\left|\frac{\left(g_{\phi}^{2}v_{\phi}\right)^{2}}{m_{h_i}^{2}-m_{V}^{2}}\right|^{2}
R_{2i}^{2} \; ,\nonumber\\ 
<\sigma v>_{AA\to h_{i}h_{i}}^{t+u}&=&\frac{1}{4\pi}\frac{1}{m_{A}^{2}}
\Bigg[\sqrt{1-\frac{m_{h_1}^{2}}{m_{A}^{2}}}
\left|\frac{\left(\delta_{2}v/2\right)^{2}R_{1i}}{m_{h_1}^{2}-m_{A}^{2}}
\right|^{2}+\sqrt{1-\frac{m_{h_2}^{2}}{m_{A}^{2}}}
\left|\frac{\left(\delta_{1}v_{\phi}/2\right)^{2}R_{2i}}{m_{h_2}^{2}-m_{A}^{2}}
\right|^{2}\nonumber\\
&+&\sqrt{1-\frac{m_{h_3}^{2}}{m_{A}^{2}}}
\left|\frac{\left(d_{2}v_{s}/2\right)^{2}R_{3i}}{m_{h_3}^{2}-m_{A}^{2}}
\right|^{2}\Bigg] \; .\nonumber
\end{eqnarray}
\item{{\bf Bottom-right}}
\begin{eqnarray}
<\sigma v>_{VV\to h_{i}h_{i}}^{seagull}&=&\frac{1}{32\pi}\frac{1}{m_{V}^{2}}
\sqrt{1-\frac{m_{h_i}^{2}}{m_{V}^{2}}}
\left|\frac{g_{\phi}^{2}}{2}\right|^{2}R_{2i}^{2}\; ,\nonumber\\
<\sigma v>_{AA\to h_{i}h_{i}}^{seagull}&=&\frac{1}{32\pi}\frac{1}{m_{A}^{2}}
\Bigg[\sqrt{1-\frac{m_{h_1}^{2}}{m_{A}^{2}}}
\left|\frac{\delta_{2}}{2}
\right|^{2}R_{1i}^{2}+\sqrt{1-\frac{m_{h_2}^{2}}{m_{A}^{2}}}
\left|\frac{\delta_{1}}{2}\frac{v_{\phi}}{v}
\right|^{2}R_{2i}^{2}\nonumber\\ 
&+&\sqrt{1-\frac{m_{h_3}^{2}}{m_{A}^{2}}}
\left|\frac{d_{2}}{2}\frac{v_{s}}{v}\right|^{2}R_{3i}^{2}\Bigg]\; .\nonumber
\end{eqnarray}

\end{itemize}
%\end{enumerate}

\section{Boltzman equations}

The coupled boltzman equations could be written as~\cite{Bian:2013wna}:
\begin{eqnarray}
\label{eq:boltzAA}
  \frac{d Y_{A}}{d x_{A}}&=&-
  \frac{1.32g^{1/2}_{\star}M_{A}M_{p}}{x^{2}_{A}}
  \bigg(\langle\sigma v_{rel}\rangle_{AA\rightarrow X\bar{X}}
  \big(Y^{2}_{A}-(Y^{eq}_{A})^{2}\big)\nonumber\\
  &+&
  \langle\sigma v_{rel}\rangle_{AA\rightarrow VV}
  \bigg(Y^{2}_{A}-\frac{(Y^{eq}_{A})^{2}}{(Y^{eq}_V)^{2}}Y^{2}_{V}\bigg)\bigg)\;,\\
\label{eq:boltzVA}
  \frac{d Y_V}{d x_V}&=&-
  \frac{1.32g^{1/2}_{\star}M_VM_{p}}{x^{2}_V}
  \bigg(\langle\sigma v_{rel}\rangle_{VV\rightarrow X\bar{X}}
  \big(Y^{2}_V-(Y^{eq}_V)^{2}\big)\nonumber\\
  &-&
  \langle\sigma v_{rel}\rangle_{AA\rightarrow VV}
  \bigg(Y^{2}_{A}-\frac{(Y^{eq}_{A})^{2}}{(Y^{eq}_V)^{2}}Y^{2}_V\bigg)\bigg)\;,
\end{eqnarray}
for $M_{A}>M_{V}$. 

Similarly, for $M_{V}>M_{A}$, one has
\begin{eqnarray}
\label{eq:boltzVV}
  \frac{d Y_V}{d x_V}&=&-
  \frac{1.32g^{1/2}_{\star}M_V M_{p}}{x^{2}_V}
  \bigg(\langle\sigma v_{rel}\rangle_{VV\rightarrow X\bar{X}}
  \big(Y^{2}_V-(Y^{eq}_V)^{2}\big)\nonumber\\
  &+& 
  \langle\sigma v_{rel}\rangle_{VV\rightarrow AA}
  \bigg(Y^{2}_{V}-\frac{(Y^{eq}_{V})^{2}}{(Y^{eq}_A)^{2}}Y^{2}_A\bigg)\bigg)\;,\\
\label{eq:boltzAV}
  \frac{d Y_{A}}{d x_{A}}&=&-
  \frac{1.32g^{1/2}_{\star}M_{A}M_{p}}{x^{2}_{A}}
  \bigg(\langle\sigma v_{rel}\rangle_{AA\rightarrow X\bar{X}}
  \big(Y^{2}_{A}-(Y^{eq}_{A})^{2}\big)\nonumber\\
  &-&
  \langle\sigma v_{rel}\rangle_{VV\rightarrow AA}
  \big(Y^{2}_{V}-\frac{(Y^{eq}_{V})^{2}}{(Y^{eq}_A)^{2}}Y^{2}_A\bigg)\bigg)\;.
\end{eqnarray}
Here, $M_{p}=2.44\times10^{18}$ GeV is the reduced Planck mass, 
and $g_{\star}$ is the degrees of freedom parameter. Two dimensionless variables $Y_{i,j}$ 
relate with number density through $Y_{i,j}=\frac{n_{i,j}}{s}$, $x_{i,j}=\frac{m_{i,j}}{T}$~\cite{Gondolo:1990dk}, 
and $s$ and $T$ are the entropy density and temperature of the Universe. After solving the coupled 
Eqs.~(\ref{eq:boltzAA}), (\ref{eq:boltzVA}), (\ref{eq:boltzVV}), and (\ref{eq:boltzAV}), 
one gets the values of $Y_{A}$ and $Y_{V}$ at present temperature $T_{0}$.

\section{One loop $\beta$ functions}

The one-loop renormalization group equations (RGEs), which we used to analyze the vacuum stability problem, are given by
\begin{eqnarray}
\label{beta}
\frac{d X}{d \log \mu}=\frac{1}{16\pi^2}\beta_X\; ,
\end{eqnarray}
with one-loop $\beta$-functions $\beta_X$,
\begin{eqnarray}
\beta_{\delta_1}&=&\frac{1}{2}(4d_2 \delta_1-9g_\phi^2\delta_1^2+4\delta_1^2+3\delta_1\lambda_\phi+2\lambda_m\delta_2)\; ,\\
\beta_{\delta_2}&=&\frac{1}{2}(4 d_2 \delta_2-3g_1^2\delta_2-9g_2^2\delta_2+12g_t^2\delta_2+4\delta_2^2
+12\delta_2\lambda+2\delta_1\lambda_m)\; ,\\
\beta_{d_2}&=&\frac{1}{2}(10 d_2^2+\delta_1^2+\delta_2^2)\; ,\\
\beta_{\lambda_m}&=&\frac{1}{2}(\delta_1 \delta_2-3 g_1^2 \lambda_m
-9 g_2^2 \lambda_m+12g_t^2\lambda_m-9g_\phi^2\lambda_m+12\lambda\lambda_m+8\lambda_m^2+3\lambda_m\lambda_\phi)\; ,\\
\beta_{\lambda}&=&-\lambda(3g^2_{1}
+9g^2_{2}-12g^{2}_{t}-24\lambda)+\frac{3}{4}g^4_{2}+
\frac{3}{8}(g^2_{1}+g^2_{2})^2-6g^4_{t}+\frac{\delta_2^2}{4}+\frac{\lambda_m^2}{2}\;,\\
\beta_{\lambda_\phi}&=&\frac{1}{2}(9 g_\phi^4+2\delta_1^2
+4\lambda_m^2-18g_\phi^2\lambda_\phi+9\lambda_\phi^2)\; ,\\
\beta_{g_\phi}&=&-\frac{43}{6}g_\phi^3 \; .
\end{eqnarray}

\acknowledgments

LGB thank Michael Ramsey-Musuolf, Ran Ding and Bin Zhu for useful discussions.
This research was supported in part by the Natural Science
Foundation of China under grant numbers 10821504, 11075194, 11135003, 11275246, and 11475238,
and by the National Basic Research Program of China (973 Program) under grant number 2010CB833000.

%%%%%%%%%%%%%%%%%%%%%%%%%%%%%%%%%%%%%%%

%%%%%%%%%%%%%%%%%%%%%%%%%%%%%%%%%%%%%%%%%%%%%%%%%%%
\end{document}